\documentclass[12pt]{article}

\usepackage{amstext,amsmath,amssymb,amsfonts,bbm}
\usepackage[latin1]{inputenc}
\usepackage{epsfig}
\usepackage{hyperref}
\usepackage{amsthm}
\usepackage{subfigure}
\usepackage{color}

\newcommand{\bea}{\begin{eqnarray}}
\newcommand{\eea}{\end{eqnarray}}

\newtheorem{remark}{Remark}

\newtheorem{theorem}{Theorem}

\makeatletter

\begin{document}

\begin{titlepage}
\begin{flushright}
pi-qg-198\\
ICMPA-MPA/2010/16
\end{flushright}

\bigskip

\begin{center}

{\Large\bf Asymptotes in $SU(2)$ Recoupling Theory:}\\
{\Large \bf Wigner Matrices, $3j$ Symbols, and Character Localization}\\

\bigskip

Joseph Ben Geloun$^{a,b,*}$, Razvan Gurau $^{a,\dag}$

\bigskip

$^{a}${\sl Perimeter Institute for Theoretical Physics\\
31, Caroline N., ON N2L 2Y5, Waterloo, Canada}\\

\medskip

$^{b}${\sl International Chair in Mathematical Physics and Applications (ICMPA--UNESCO Chair),\\
University of Abomey--Calavi, 072 B. P. 50, Cotonou, Republic of Benin}

\bigskip

E-mail:  $^{*}${\em jbengeloun@perimeterinsitute.ca},\quad $^{\dag}${\em rgurau@perimeterinsitute.ca},

\bigskip

\today

\medskip

\begin{abstract}
\noindent
In this paper we employ a novel technique combining the Euler Maclaurin formula with 
the saddle point approximation method to obtain the asymptotic behavior 
(in the limit of large representation index $J$) of generic Wigner matrix elements 
$D^{J}_{MM'}(g)$. We use this result to derive asymptotic formulae 
for the character $\chi^J(g)$ of an $SU(2)$ group element and for Wigner's 
$3j$ symbol. Surprisingly, given that we perform five successive layers of 
approximations, the asymptotic formula we obtain for $\chi^J(g)$ is in fact 
{\it exact}. This result provides a non trivial example of a Duistermaat-Heckman like 
localization property for discrete sums.
\end{abstract}

\end{center}

\noindent Pacs: 02.20.Qs, 05.45.Mt

\end{titlepage}

\setcounter{footnote}{0}

\section{Introduction}
\label{Intro}

The saddle point approximation (SPA) is a classical algorithm to determine 
asymptotic behavior of a large class of integrals in some large parameter 
limit \cite{horman}. One uses it when exact calculations are either too 
complex or not very relevant. Recently SPA has been used in conjunction with the 
Euler Maclaurin (EM) formula to derive asymptotic behavior of discrete 
sums \cite{GU1,GU2}. In the combined EM SPA scheme corrections to the leading 
behavior come from two sources: the derivative terms in the EM formula
and sub leading terms in the SPA estimate.

In this paper we use the EM SPA method to derive the asymptotic behavior of Wigner rotation 
matrix elements. We subsequently use this asymptotic formula to derive the asymptotic 
behavior of the character of an $SU(2)$ group element. Although our estimate is obtained after 
using twice the EM SPA approximation and once the Stirling approximation
for Euler's Gamma functions it turns out to be the exact result. We then proceed to 
obtain the asymptotic expression for Wigner's $3j$ symbol, recovering with this 
method the results of \cite{Wi3j}. 

Our results are relevant for computing
topological (Turaev Viro like \cite{TV}) invariants and in connection to the volume
conjecture \cite{malek2}. From a theoretical physics perspective they are 
of consequence for spin foam models \cite{Baez:1999sr},
Group Field Theory \cite{Freidel:2005qe, Oriti:2006se}, discretized BF theory and 
lattice gravity \cite{Barrett:1997gw},\cite{Engle:2007wy}, \cite{Freidel:2007py}. 
Continuous SPA has been extensively used in this context to derive asymptotic 
behaviors of spin foam amplitudes
\cite{Barrett:2009gg}, \cite{Dowdall:2009eg}, \cite{Dupuis:2009qw}, 
and \cite{Alesci:2008ff}, \cite{Conrady:2008mk}, \cite{Krajewski:2010yq}.

In the recoupling theory of $SU(2)$, the EM SPA method has already been used to 
obtain in a particularly simple way the Ponzano-Regge asymptotic of the $6j$ symbol
 \cite{GU2}, \cite{Dupuis:2009sz}. The main strength of this approach is the following.
Most relevant quantities in the recoupling theory of $SU(2)$ are
expressed in Fourier space by discrete sums.
In particular, the Wigner matrix elements admit a single sum 
representation \cite{messiah}. However, generically, 
the sums are alternated hence difficult to handle. 
Our EM SPA method deals very efficiently with alternating signs: 
generically such signs lead to complex saddle points situated outside the initial
 summation interval. After exchanging the original sums 
(via the EM formula) for integrals one only deforms the integration contour in the complex plane
to pass trough the saddle points in a completely standard manner. This feature is the crucial strength of 
our method, and allows rapid access to explicit results.
The EM SPA method should allow one to prove for instance the asymptotic behavior \cite{Wi9j} of
the $9j$ symbol.

The proofs of our three main results (Theorems \ref{theo1}, \ref{theo2} and \ref{th:3j})
are straightforward, but the shear amount of computations performed renders this a 
somewhat technical paper. In Section 
\ref{sect2} we give a quick review of iterated saddle point approximations. In Section \ref{sect3} we establish 
Theorem \ref{theo1} and use it in Section \ref{sect4} to derive the character formula (Theorem \ref{theo2}). 
Section \ref{sec:3j} proves the asymptotic  formulae of the $3j$ symbol (Theorem \ref{th:3j}). 
Section \ref{concl} draws the conclusion of our work and discusses the relation between our 
result for the character and the Duistermaat Heckman theorem. The (very detailed) Appendices present
explicit computations and detail the EM derivative terms.

\section{Successive saddle point approximations}
\label{sect2}

We briefly review the iterated SPA approximations. The result of this section
justifies the use of our asymptote of the Wigner matrices to derive the asymptotic 
behavior of $SU(2)$ characters and Wigner $3j$ symbols. 

Consider a function $f$ of two real variables. We are interested in evaluating the 
asymptotic behavior of the integral 
\bea
I=  \int dudx \; e^{ J  f(u,x)},
\eea
for large $J$. One can chose to either evaluate $I$ via an SPA in both variables at the same time or 
via two successive SPA, one for each variable. The question is if the two estimates coincide. This problem is 
addressed in full detail in \cite{horman} and the answer to the above question
is yes (for sufficiently smooth functions), with known estimates. Let us give a quick flavor 
of the origin of this result.
\begin{remark}
\label{prop1}
Let $f:\mathbb{R} \times \mathbb{R} \rightarrow \mathbb{C}$ be a function with 
and unique critical point $(u_c,x_c)$ and non degenerate Hessian at $(u_c,x_c)$
such that $I = \int e^{ J f(u,x)}$ admits a SPA at large $J$.
Assume that the equation $\partial_u f(u,x) = 0$ admits an unique 
solution $u_c=h(x)$, such that $[\partial^2_{u}f](h(x),x)\neq 0$.
Then the SPA of $\int e^{ J f(u,x)}$  in both variables $(u,x)$
gives the same estimate as two successive SPAs, the first one in $u$ 
and the second one in $x$.

\end{remark}
\noindent{\bf Proof:} The simultaneous SPA in $u$ and $x$ yields the estimate
\bea
 I \approx \frac{2\pi}{ J \sqrt{    \bigl[\partial_u^2 f \partial_x^2 f - (\partial_u\partial_x f)^2\bigr] 
\Big{|}_{(u_c,x_c)}  } }
 e^{J \, f(u_c,x_c)} \; .
\label{eq:simulspa}
\eea
The saddle point equation for $u$, $\bigl[\partial_u f \bigr] (u,x)=0$, is solved by $u_c = h(x)$.
Thus a first SPA in $u$ yields 
\bea\label{eq:interm}
 I \approx \sqrt{\frac{2\pi}{J}} \int dx  \frac{1}{\sqrt{ - \partial^2_u f \big{|}_{(h(x),x)} }}
 e^{J \, f(h(x),\, x)}.
\eea
We evaluate eq. (\ref{eq:interm}) by a second SPA, in the $x$ variable. The saddle point equation is
\bea
\frac{d}{dx} \Big{(} f\bigl(h(x),x \bigr) \Big{)}=
[\partial_u f] \Big{|}_{(h(x),x)}\; \frac{dh}{dx} + [\partial_x f]\Big{|}_{(h(x),x)} \; ,
\eea
and, as $[\partial_u f](h(x),x) =0 $, the first term 
above vanishes. The critical point $x_c$ is therefore solution of $[\partial_x f]\Big{|}_{(h(x),x)}=0$. 
The second derivative of $f(h(x),x)$ computes to
\bea
\frac{d^2}{dx^2} \Big{(} f(h(x),x) \Big{)} = \frac{d}{dx} \Big{(}
[\partial_x f]\Big{|}_{(h(x),x)} \Big{)}
= [\partial_u\partial_x f]\Big{|}_{(h(x),x)}\frac{dh}{dx} + [\partial_x^2 f]\Big{|}_{(h(x),x)} \; ,
\eea
and noting that
\begin{equation}
\frac{d}{dx} \bigl[\partial_u f\bigr]\Big{|}_{(h(x),x)}  =0\Rightarrow
[\partial_u^2 f] \Big{|}_{(h(x),x)} \frac{dh}{dx}+ \partial_x [\partial_u f] \Big{|}_{(h(x),x)} =0 \Rightarrow
 \frac{dh}{dx} = -\frac{ [\partial_x \partial_u f]}{[\partial_u^2 f]} \Big{|}_{(h(x),x)},
\end{equation}
the estimate obtained by two successive SPAs is 
\bea
 I \approx
 \frac{2\pi}{J \,\sqrt{   \partial^2_u f \Big{|}_{(h(x_c),x_c)}
\Big{(}
-\frac{[\partial_u\partial_x f]^2}{[\partial_u^2 f]} + \partial_x^2 f \Big{)} \Big{|}_{(h(x_c),x_c)} } }
e^{J \, f(u_c, \,x_c)} \; ,
\eea
identical with eq. (\ref{eq:simulspa}).

\qed

This remark generalizes \cite{horman}, for sufficiently smooth functions of more 
variables with non degenerate critical points. In the sequel we will express the 
Wigner matrix elements $D^J_{MM'}$ (up to corrections coming from the EM formula) as 
integrals which we approximate by a first SPA. To compute more involved sums or integrals 
of products of such matrix elements (the character of a $SU(2)$ group element and 
the $3j$ symbol) we will substitute the SPA approximation for each $D^J_{MM'}$
and evaluate the resulting expressions by subsequent SPAs.

\section{Asymptotic formula of a Wigner matrix element}
\label{sect3}

In this section we prove an asymptotic formula for a 
Wigner matrix element. Before proceeding let us mention that many of our results 
write in terms of angles. In order to avoid issues related to the interval of definition 
of this angles we will always denote them as $\imath\phi=\ln w$ 
for some complex number $w$ with $|w|=1$.

Our starting point is the classical expression of
$D^J_{MM'}$ in terms of Euler angles $(\alpha,\beta,\gamma)$ in $z\; y\;z$ order 
(see \cite{messiah})
\bea
 D^{J}_{MM'}(\alpha,\beta,\gamma) =&& e^{-\imath\alpha M} e^{-\imath \gamma M'} \sum_{t} (-)^t
\frac{\sqrt{(J+M)!(J-M)!(J+M')!(J-M')!}}{(J+M-t)!(J-M'-t)!t!(t-M+M')!}
\crcr
&& \xi^{2J+M-M'-2t} \eta^{2t-M+M'} \; ,
\label{eq:primdj}
\eea
with $\xi=\cos (\beta/2), \;\eta=\sin(\beta/2)$.
The sum is taken over all $t$ such that
all factorials have positive argument (hence it has 
$1+ \min \{J+M,J-M,J+M',J-M' \}$ terms). We call a Wigner 
matrix generic if its second Euler angle $\beta \notin \mathbb{Z} \pi $ (that is $0<\xi^2<1$).
We define the reduced variables $x= \frac{J}{M}$ and $y= \frac{J}{M'}$. A priori the asymptotic
behavior we derive below holds in certain region of the parameters $x$, $y$ and $\xi$ 
detailed in Appendices \ref{app:EM} and \ref{app:delta<0}. 
\begin{theorem}
\label{theo1}
A generic Wigner matrix element in the spin $J$ representation of a $SU(2)$ 
group element has in the large $J$ limit the asymptotic behavior
 \bea
\label{eq:asymp}
D^{J}_{xJ,yJ}(\alpha,\beta,\gamma)
&\approx& e^{-\imath J \alpha x - \imath J \gamma y}
  \Big{(} \frac{1} {  \pi J  \sqrt{\Delta}}\Big{)}^{\frac{1}{2}} 
\cos\Big{[} \Bigl(J+\frac{1}{2}\Bigr) \phi + x J \psi - y J \omega  -\frac{\pi}{4}\Big{]} \; ,
\eea
with
\bea
 \Delta = (1-\xi^2)(\xi^2 -xy) - \frac{(x-y)^2}{4}\geq 0 \; ,
\eea
with $ \phi, \psi $ and $ \omega $ the three angles 
\bea\label{eq:angles}
&&\imath \phi =
\ln \frac{2\xi^2 -1-xy +2\imath \sqrt{\Delta}}{ \sqrt{(1-x^2)(1-y^2)}} \;,
\quad
\imath \psi 
=\ln \frac{ \frac{x+y}{2} - x\xi^2 + \imath\sqrt{\Delta} }{\sqrt{\xi^2(1-\xi^2)(1-x^2)}} \;,
\nonumber\\
&& \imath \omega  =\ln \frac{-\frac{x+y}{2} + y \xi^2
+ \imath  \sqrt{\Delta} }
{\sqrt{\xi^2(1-\xi^2)(1-y^2)} } \; .
\eea
\end{theorem}

\noindent{\bf Proof:}
The proof of Theorem \ref{theo1} is divided into two steps: first the approximation of
eq. (\ref{eq:primdj}) by an integral via the EM formula, and second the evaluation 
of the latter by an SPA.

\noindent{\bf Step 1:} In the large $J$ limit the leading behavior of the Wigner matrix element eq. (\ref{eq:primdj}) is
\bea
D^{J}_{xJ,yJ}(\alpha,\beta,\gamma) \approx \frac{1}{2\pi } \int du\,
\sqrt{K(x,y,u)}\,
e^{J f(x,y,u)} \; ,
\label{eq:dlam}
\eea
where
\bea
&&f(x,y,u) =
-\imath\alpha x -\imath \gamma y +\imath \pi u + (2+x-y-2u) \ln \xi
+ (2u-x+y) \ln \eta  \crcr
&&
+ \frac{1}{2}(1-x) \ln(1-x) +  \frac{1}{2}(1+x) \ln(1+x)
 +  \frac{1}{2}(1-y) \ln(1-y) + \frac{1}{2}(1+y) \ln(1+y) \crcr
&&-  (1+x-u)\ln(1+x-u) - (1-y-u) \ln(1-y-u)\crcr
&& -u\ln u - (u-x+y) \ln(u-x+y) \; ,
\label{eq:primf}
\eea
and 
\bea\label{eq:primK}
 K(x,y,u) = \frac{\sqrt{(1-x)(1+x)(1-y)(1+y)} }
{(1+x-u)(1-y-u)(u)(u-x+y) } \; .
\eea
To prove this we rewrite eq. (\ref{eq:primdj}) in terms of Gamma functions 
\bea
&& D^{J}_{MM'}(\alpha,\beta,\gamma) = \sum_{t} F(J,M,M',t) \; , \nonumber\\
&& F(J,M,M',t) = e^{\imath\pi t} e^{-\imath\alpha M} e^{-\imath \gamma M'} \xi^{2J+M-M'-2t} \eta^{2t-M+M'} \times
\nonumber\\
&&\frac{\sqrt{\Gamma(J+M+1)\Gamma(J-M+1)\Gamma(J+M'+1)\Gamma(J-M'+1)}}{ \Gamma(J+M-t+1)
\Gamma(J-M'-t+1)\Gamma(t+1)\Gamma(t-M+M'+1)} \; ,
\label{eq:djmm}
\eea
and use the Euler-Maclaurin
formula 
\bea
\sum_{t_{\min}}^{t_{\max}} h(t) &=&\int_{t_{\min}}^{t_{\max}} h(t) dt - B_1\bigl[h(t_{\max})+h(t_{\min})\bigr] \nonumber\\
&+& \sum_k \frac{B_{2k}}{(2k)!} \bigl[h^{(2k-1)}(t_{\max}) - h^{(2k-1)}(t_{\min}) \bigr] \;,
\label{eq:eumac}
\eea
where $B_1,B_{2k}$ are the Bernoulli numbers\footnote{Eq. (\ref{eq:eumac}) holds for all $C^{\infty} $ 
functions $h(t)$, such that the sum over $k$ converges. }. 
To derive our asymptote we only take into account the integral approximation of eq. (\ref{eq:djmm})
(the boundary terms are discussed in Appendix \ref{app:EM}), hence
\bea\label{eq:DfromF}
D^{J}_{MM'}(\alpha,\beta,\gamma) \approx \int dt\; F(J,M,M',t) \; .
\eea

We define $u=\frac{t}{J}$ hence $du=\frac{1}{J} dt$ and using the Stirling formula for the Gamma functions 
(see Appendix \ref{app:stir}) we get eq. (\ref{eq:dlam}).

\noindent{\bf Step 2:} We now proceed to evaluate the integral (\ref{eq:dlam}) by an SPA. 
Some of the computations relevant for this proof are included in Appendix \ref{app:eval}.
Denoting the set of saddle points by $\mathcal{C}$, the leading asymptotic behavior of a
 generic Wigner matrix element writes
\bea\label{eq:Dsaddle}
 D^J_{xJ,yJ} (\alpha,\beta,\gamma)\approx 
\frac{1}{\sqrt{2\pi J}} \,
\sum_{u_* \in \mathcal{C}} \,
 \frac{ \sqrt{K{|}_{x,y,u_*}}}{\sqrt{ (-\partial_u^2 f)|_{x,y,u_*} }}\;
e^{J f(x,y,u_*)}.
\eea
Our task is to identify $\mathcal{C}$ and compute $K{|}_{x,y,u_*} $, $ (-\partial_u^2 f)|_{x,y,u_*} $ and $f(x,y,u_*)$.

\medskip

\noindent{\bf The set $\mathcal{C}$.} The derivative of $f$ with respect to $u$ is
\bea\label{eq:deriv}
 \partial_u f = \imath \pi -2  \ln \xi + 2 \ln \eta + \ln(1+x-u) + \ln(1-y-u)- \ln u - \ln(u-x+y) \; .
\eea
A straightforward computation shows that the saddle points are the solutions of
\bea
&& (1+x-u) (1-y-u) \frac{(1-\xi^2)}{\xi^2} + u (u-x+y) =0 \label{eq:SPE}\\
&&\Leftrightarrow
u^{2} - u [ 2 (1-\xi^2) +x-y   ] + (1-\xi^2) (1+x)(1-y)=0 \; .
\label{eq:SPE2}
\eea

The region of parameters $x,y,\xi$ for which the discriminant of 
eq. (\ref{eq:SPE2}) is positive gives exponentially suppressed matrix elements while 
the region for which it is zero gives an Airy function estimate. Both cases are detailed in 
Appendix \ref{app:delta<0}. 

In the rest of this proof we treat the region in which the discriminant of eq. (\ref{eq:SPE2}) is 
negative. We denote by $\Delta$ minus the reduced discriminant, that is
\bea
 \Delta = (1-\xi^2)(\xi^2-xy) - \frac{(x-y)^2}{4} > 0 \; ,
\label{eq:discriminant}
\eea
and the two saddle points, solutions of eq. (\ref{eq:SPE2}), write
\bea
 u_{\pm} = (1-\xi^2) + \frac{x-y}{2}
 \pm \imath \sqrt{\Delta} \; ,
\eea
thus the set of saddle points is $\mathcal{C}=\{u_+,u_-\}$.

\medskip

\noindent{\bf Evaluation of $f(x,y,u_{\pm})$.} We rearrange the terms in 
eq. (\ref{eq:primf}) to write
\bea
&&
f(x,y,u) = -\imath\alpha x -\imath \gamma y + (2+x-y) \ln \xi
+ (-x+y) \ln \eta \nonumber\\
&&+ \frac{1}{2}(1-x) \ln(1-x) +  \frac{1}{2}(1+x) \ln(1+x)
 +  \frac{1}{2}(1-y) \ln(1-y) + \frac{1}{2}(1+y) \ln(1+y) \nonumber\\
&&-  (1+x)\ln(1+x-u) - (1-y) \ln(1-y-u)  - (-x+y) \ln(u-x+y)
\nonumber\\
&& + u \ln\Big{[} (-) \frac{1-\xi^2}{\xi^2} \; \frac{(1+x-u)(1-y-u)}{u (u-x+y)} \Big{]} \;.
\label{eq:reducedf}
\eea

Note that by the saddle point equations the last line in eq. (\ref{eq:reducedf}) is zero for $u_{\pm}$.
The rest of eq. (\ref{eq:reducedf}) computes to (see Appendix \ref{app:critf} for details) 
\bea
 f(x,y,u_{\pm}) = -\imath\alpha x -\imath \gamma y \pm \imath \Big{(} \phi + x\psi - y\omega \Big{)} \; ,
\label{eq:ffin}
\eea
with 
\bea
&&\imath \phi =
\ln \frac{2\xi^2 -1-xy +2\imath \sqrt{\Delta}}{ \sqrt{(1-x^2)(1-y^2)}} \;,
\quad
\imath \psi 
=\ln \frac{ \frac{x+y}{2} - x\xi^2 + \imath\sqrt{\Delta} }{\sqrt{\xi^2(1-\xi^2)(1-x^2)}} \;,
\nonumber\\
&& \imath \omega  =\ln \frac{-\frac{x+y}{2} + y \xi^2
+ \imath  \sqrt{\Delta} }
{\sqrt{\xi^2(1-\xi^2)(1-y^2)} } \; .
\label{eq:phichiom}
\eea

\medskip

\noindent{\bf Second derivative.} The derivative of eq. (\ref{eq:deriv}) is 
\bea
 -\partial_u^2 f(x,y,u) = \frac{1}{1+x-u} +\frac{1}{1-y-u} +\frac{1}{u} +\frac{1}{u-x+y} \; .
\label{eq:uhess}
\eea
At the saddle points a straightforward computation shows that the second derivative is (see Appendix \ref{app:critderiv})
\bea
 (-\partial_u^2 f) \Big{|}_{x,y,u_\pm}  = \frac{1}{(1-x^2)(1-y^2) \xi^2 (1-\xi^2)} \Big{(} 4\Delta  \pm  \imath 2 \sqrt{\Delta}
\bigr[1+ xy - 2  \xi^2  \bigl]\Big{)} \;.
\label{eq:finsec}
\eea

\medskip

\noindent{\bf The prefactor $K$.} The prefactor $K(x,y,u)$ is 
\bea
  K  = \frac{ \sqrt{(1-x^2)(1-y^2) } }
 {u(1+x-u)(1-y-u)(u-x+y) } \; ,
\label{eq:pref}
\eea
which computes at the saddle points to (see Appendix \ref{app:critK})
\bea
K \Big{|}_{x,y,u_\pm} = \frac{-\sqrt{(1-x^2)(1-y^2) }
\Big{(} 2 \xi^2-1-xy \pm 2\imath \sqrt \Delta\Big{)}^2}{\xi^2 (1-\xi^2) (1-x^2)^2(1-y^2)^2} \;.
\label{eq:kfac}
\eea

\medskip

\noindent{\bf Final evaluation.} Before collecting all our previous results we first evaluate, using 
eq. (\ref{eq:finsec}) and (\ref{eq:kfac}) 
\bea\label{eq:raport}
\frac{K|_{x,y,u_{\pm}}}{(-\partial_u^2 f)|_{x,y,u_{\pm}}} 
&=& -\frac{\Big{(} 2 \xi^2-1-xy \pm 2\imath \sqrt \Delta\Big{)}^2}{\sqrt{(1-x^2)(1-y^2) }
\Big{(} 4\Delta  \pm  \imath 2 \sqrt{\Delta}
\bigr[1+ xy - 2  \xi^2  \bigl] \Big{)} } \nonumber\\
&=& \frac{1}{\sqrt{(1-x^2)(1-y^2) } ( \pm 2\imath\sqrt{\Delta} ) }
\Big{(} 2 \xi^2-1-xy \pm 2\imath \sqrt \Delta\Big{)} \nonumber\\
&=& \frac{1}{\pm \imath 2\sqrt{\Delta}}\frac{ \Big{(} 2 \xi^2-1-xy \pm 2\imath \sqrt \Delta\Big{)}}{ \sqrt{(1-x^2)(1-y^2) }}
\; .
\eea
Comparing eq. (\ref{eq:raport}) with eq. (\ref{eq:angles}) we conclude that 
\bea\label{eq:preffin}
\frac{K|_{x,y,u_\pm}}{(-\partial_u^2 f)|_{x,y,u_{\pm}}} 
= \frac{1}{\pm \imath2\sqrt{\Delta}}e^{\pm \imath \phi} \; .
\eea
Substituting eq. (\ref{eq:preffin}) and (\ref{eq:ffin}) into eq. (\ref{eq:Dsaddle})
we obtain
\bea\label{eq:asymp2}
 D^J_{xJ,yJ} (\alpha,\beta,\gamma) &\approx&
\frac{1}{\sqrt{2\pi J}} \Big{(} \frac{1} {2 \sqrt{\Delta}}\Big{)}^{\frac{1}{2}} 
e^{-\imath J \alpha x - \imath J \gamma y}
\crcr
&& \Big{(} \sqrt{\frac{1}{\imath}e^{\imath \phi} }  \; e^{\imath J ( \phi
+ x \psi - y \omega) } +
\sqrt{\frac{1}{-\imath} e^{-\imath \phi} } \; e^{ - \imath J ( \phi
+ x \psi - y \omega) }
 \Big{)} \; ,
\eea
and a straightforward computation proves Theorem \ref{theo1}. 

\qed

\section{Characters}
\label{sect4}

In this section we use Theorem \ref{theo1} to derive an asymptotic formula for the character of
a $SU(2)$ group element. 
\begin{theorem}\label{theo2}
The leading asymptotic behavior of the character of a $SU(2)$ group element 
(with Euler angles $(\alpha,\beta,\gamma)$) in the $J$ representation, $\chi^J(\alpha,\beta,\gamma)$ is
\bea\label{eq:theo2}
 \chi^J(\alpha,\beta,\gamma) \approx  \
\frac{\sin \Big{[} \bigl(J+\frac{1}{2} \bigr)\theta\Big{]}}{\sin\frac{\theta}{2}} \; ,
\eea
with $\theta$ defined by 
\bea\label{eq:theo2thet}
\cos \frac{\theta}{2} = \cos\frac{\beta}{2}\,  \cos\frac{(\alpha + \gamma) }{2} \;.
\eea
\end{theorem}

Let us emphasize that up to this point we already performed three 
different approximations: first the EM approximation, second the Stirling approximation 
and third the SPA approximation. To prove Theorem \ref{theo2} we will use a second EM 
approximation and a second SPA approximation. However, formula (\ref{eq:theo2thet})
is exactly the classical relation between the Euler angle parameterization and the 
$\theta,\vec n$ parameterization of an $SU(2)$ group element, thus the leading behavior
we find (after five levels of approximation) is in fact the exact
formula of the character! We will discuss this rather surprising
result in Section \ref{concl}.

\medskip

\noindent{\bf Proof of Theorem \ref{theo2}.} To establish Theorem \ref{theo2} we follow again the
EM SPA recipe. The character $\chi^J$ of a group element writes 
\bea\label{eq:charsum}
\chi^J(\alpha,\beta,\gamma) = \sum^J_{M=-J} D^J_{MM}(\alpha,\beta,\gamma) = \sum^1_{x=-1}
 D^{J}_{x J,x J }(\alpha,\beta,\gamma) \; ,
\eea
with $ x= \frac{M}{J}$ the rescaled variable. Note that the step in the second sum is $dx = \frac{1}{J}$.
The leading EM approximation (see end of Appendix \ref{app:EM}) for the character is therefore the 
continuous integral (dropping henceforth the argument $(\alpha,\beta,\gamma)$)
\bea\label{eq:char}
 \chi^{J} \approx J \int_{-1}^1 dx \; D^{J}_{x J,x J} \; .
\eea
We now use Theorem \ref{theo1} (more precisely eq. (\ref{eq:asymp2})) and write a diagonal
Wigner matrix element as 
\bea\label{eq:ddiag}
 D^{J}_{x J,x J} \approx   \Big{[} \frac{1} {4 \pi J \sqrt{\Delta}}\Big{]}^{\frac{1}{2}}
\Big{[} \sqrt{\frac{ e^{\imath\phi}}{\imath}}\,  e^{ J f(x,x,u_+) }
+ \sqrt{\frac{ e^{-\imath\phi}}{-\imath}} \, e^{ J f(x,x,u_-) }
 \Big{]}  \; .
\eea
Note that for diagonal matrix elements the exponents simplify to 
\bea\label{eq:funcmese}
f(x,x,u_\pm) &=& -\imath (\alpha+\gamma) x \pm \imath \Big{(} \phi +   x (\psi-\omega) \Big{)} \; , 
\eea
while the discriminant $\Delta$ and angles $\phi$, $\psi$ and $\omega$ from eq. (\ref{eq:angles}) become
\bea
&& \imath \phi = \ln \frac{2 \xi^2-1-x^2 + 2 \imath \sqrt{\Delta}}{(1-x^2)} \; ,\qquad
 \imath \psi =
\ln \frac{ x (1 - \xi^2) + \imath\sqrt{\Delta} }{\sqrt{\xi^2(1-\xi^2)(1-x^2)}},\\
&&
\imath \omega =\ln
\frac{-x (1 - \xi^2)  + \imath  \sqrt{\Delta}}{\sqrt{\xi^2(1-\xi^2)(1-x^2)} } \;, \qquad
\Delta = (1-\xi^2)(\xi^2 -x^2) \;.
\eea

We follow the same steps as in the proof of Theorem \ref{theo1}.

\noindent{\bf Critical set $\mathcal{C}_\chi$.}
The derivatives of the exponents for each of the two terms in eq. (\ref{eq:ddiag}) are
\bea
\partial_{x} f(x,x,u_{\pm}) &=&
-\imath(\alpha+\gamma) \pm  \imath (\psi-\omega)  \pm \imath \partial_x \phi 
\pm \imath x \partial_x (\psi - \omega) \; . 
\label{eq:SPE3}
\eea
The derivative of $\phi$ computes to 
\bea\label{eq:partphi}
 \imath \partial_{x}\phi = \partial_x \Bigl[ \ln ( \sqrt{\xi^2 -x^2}
 + \imath \sqrt{1-\xi^2})^2 -\ln (1-x^2) \Bigr]
= \imath \frac{2x \sqrt{1-\xi^2}} {(1-x^2)\sqrt{\xi^2 -x^2}} \; .
\eea
The difference $\psi-\omega$ computes to 
\bea
\imath (\psi-\omega) = \ln \frac{ x (1 - \xi^2) + \imath\sqrt{\Delta}}{ - x (1 - \xi^2)
+ \imath\sqrt{\Delta}}
=\ln \frac{ \Bigl( \sqrt{\xi^2-x^2}  - \imath x \sqrt{1-\xi^2} \Bigr)^2  }{ \xi^2(1-x^2)} \; ,
\eea
and its derivative 
\bea\label{eq:partpsiomeg}
\imath \partial_x (\psi-\omega) = 2 \frac{\frac{-x}{\sqrt{\xi^2-x^2}} - \imath \sqrt{1-\xi^2} }
{\sqrt{\xi^2-x^2} - \imath x \sqrt{1-\xi^2}} - \frac{-2x}{1-x^2}
= \imath \frac{ - 2 \sqrt{1-\xi^2}}{(1-x^2)\sqrt{\xi^2 - x^2}} \;.
\eea
Combining eq. (\ref{eq:partphi}) and (\ref{eq:partpsiomeg}) we have
\bea\label{eq:smeche}
 \partial_x\phi  +x \partial_x (\psi-\omega) = 0 \;,
\eea
and the saddle point equations (\ref{eq:SPE3}) simplify to
\bea\label{eq:smeche1}
 \psi-\omega = \pm (\alpha+\gamma) \; .
\eea
Dividing by 2 and exponentiating we get 
\bea\label{eq:tan}
  \frac{\sqrt{\xi^2-x^2}  - \imath x \sqrt{1-\xi^2}}{\sqrt{\xi^2(1-x^2)}}
   = e^{\pm \imath \frac{\alpha+\gamma}{2}} 
\Rightarrow \frac{x \sqrt{1-\xi^2}}{\sqrt{\xi^2-x^2} } = \mp \tan \frac{\alpha+\gamma}{2}  \;.
\eea
Hence the saddle points are solutions of the quadratic equation
\bea
x^2(1-\xi^2)= (\xi^2-x^2) \tan^2\frac{\alpha+\gamma}{2} \Rightarrow  x^2 =
 \frac{\xi^2 \sin^2\frac{\alpha+\gamma}{2}}{1- \xi^2\cos^2\frac{\alpha+\gamma}{2}} \; .
\eea
Defining a new variable $\theta$ via the relation $\cos\frac{\theta}{2}= \xi \cos\frac{\alpha+\gamma}{2}$ the saddle points rewrite
\bea
 x^2 =  \frac{\xi^2 \sin^2\frac{\alpha+\gamma}{2}} {\sin^2\frac{\theta}{2}} \; .
\eea
Taking into account eq. (\ref{eq:tan}) one identifies an unique saddle point ($x_1$) for $f(x,x,u_+)$ 
and an unique saddle point ($x_2$) for $f(x,x,u_-)$ with $x_1$ and $x_2$ given by
\bea
 x_1= - \frac{\xi \sin\frac{\alpha+\gamma}{2}}{ \sin\frac{\theta}{2}} \; ,\qquad
 x_2=  \frac{\xi \sin\frac{\alpha+\gamma}{2}}{ \sin\frac{\theta}{2}} \; .
\eea

\medskip

\noindent{\bf Evaluation of the functions and Hessian on $\mathcal{C}_\chi$.}
Straightforward computations lead to 
\bea
 \xi^2-x_{1,2}^2 =
(1-\xi^2) \frac{\cos^2\frac{\theta}{2}}{\sin^2\frac{\theta}{2}} \; ,\qquad
\Delta |_{x_{1,2}} = (1-\xi^2)^2 \frac{\cos^2\frac{\theta}{2}}{\sin^2\frac{\theta}{2}}   \geq 0 \; ,
\qquad  1-x_{1,2}^2 =\frac{(1-\xi^2)}{\sin^2\frac{\theta}{2}} \; .
\eea
Also note that at the saddle points $\phi$ simplifies as
\bea
&& \imath \phi = \ln \frac{2 \xi^2-1-x_{1,2}^2 + 2 \imath \sqrt{\Delta{|}_{x_{1,2}}}} {(1-x_{1,2}^2)}
= \ln \frac{ (1-\xi^2) \frac{\cos^2\frac{\theta}{2}}{\sin^2\frac{\theta}{2}}
- (1-\xi^2)+2\imath  (1-\xi^2) \frac{\cos\frac{\theta}{2}}{\sin\frac{\theta}{2}} }{\frac{(1-\xi^2)}{\sin^2\frac{\theta}{2}}  }
\nonumber\\
&&=\ln \Big{[} \cos^2\frac{\theta}{2}-\sin^2\frac{\theta}{2}
 +\imath 2\cos\frac{\theta}{2}\sin\frac{\theta}{2}  \Big{]} =\ln e^{i\theta} = i\theta \;.
\eea
Substituting the saddle point equations (\ref{eq:smeche1}) into eq. (\ref{eq:funcmese}), we see that,
at the saddles   
\bea\label{eq:funcfin}
 f(x_1,x_1,u_+)= \imath \phi = \imath\theta \; ,\qquad f(x_2,x_2,u_-)= - \imath \phi=  -\imath\theta \; .
\eea
To evaluate the Hessian at the saddle we first simplify eq. (\ref{eq:SPE3}) using eq. (\ref{eq:smeche}) hence 
\bea
\partial_x^2 f(x,x,u_{\pm})  &=&
 \pm \imath \partial_x(\psi-\omega) =
 \mp 2\imath\frac{\sqrt{1-\xi^2}}{(1-x^2)\sqrt{\xi^2-x^2}}
\eea
which becomes at the saddle points
\bea\label{eq:derivfin}
 \mp 2 \imath \frac{\sqrt{1-\xi^2}}{ \frac{(1-\xi^2)}{\sin^2\frac{\theta}{2}}
\sqrt{(1-\xi^2)} \frac{\cos\frac{\theta}{2}}{\sin\frac{\theta}{2}}
 } = \mp 2 \imath\frac{1}{1-\xi^2} \frac{\sin^3\frac{\theta}{2}}{\cos\frac{\theta}{2}} \; .
\eea

\medskip

\noindent{\bf Final evaluation.} Using eq. (\ref{eq:funcfin}) and eq. (\ref{eq:derivfin}), the SPA of
the character eq. (\ref{eq:char}) is
\bea
 \chi^J \approx \frac{1}{\sqrt{ 2 (1-\xi^2) \frac{\cos\frac{\theta}{2}}{\sin\frac{\theta}{2}} }}
  \Big{(} \sqrt{\frac{ e^{\imath \theta}} {\imath} } \frac{e^{\imath J \theta }}
{ \sqrt{ \imath \frac{2}{1-\xi^2} \frac{\sin^3\frac{\theta}{2}}{\cos\frac{\theta}{2}}}}
 + \sqrt{\frac{ e^{-\imath \theta}} {-\imath} } \frac{e^{-\imath J \theta }}
{ \sqrt{ -\imath \frac{2}{1-\xi^2} \frac{\sin^3\frac{\theta}{2}}{\cos\frac{\theta}{2}}}}
\Big{)} \; ,
\eea
which is
\bea
 \chi^J \approx \frac{1}{2 \sin\frac{\theta}{2}} \Big{(} \frac{1}{\imath }
 e^{\imath (J+\frac{1}{2})\theta}+ \frac{1}{-\imath}
 e^{-\imath(J+\frac{1}{2})\theta} \Big{)}
  = \frac{\sin \Bigl[ (J+\frac{1}{2})\theta \Bigr]}{\sin\frac{\theta}{2}}
\eea

\qed

\section{Asymptotes of $3j$ symbols}
\label{sec:3j}

In this section we employ the asymptotic formula for the Wigner matrices to obtain an 
asymptotic formula for Wigner's $3j$ symbol. Note that one can use directly the EM SPA
method to derive this asymptotic starting from the single sum representation
of the $3j$ symbol \cite{messiah}. We take here the alternative route of using
the results of Theorem \ref{theo1} and the representation of $3j$ symbols in terms of Wigner matrices
\bea
\int dg\; D^{J_1}_{M_1M_1'}(g)D^{J_1}_{M_2M_2'}(g)D^{J_3}_{M_3M_3'}(g)
= \left(\begin{array}{ccc}
 J_1&J_2&J_3\\
M_1&M_2&M_3
\end{array}\right)
\left(\begin{array}{ccc}
 J_1&J_2&J_3\\
M_1'&M_2'&M_3'
\end{array}\right)\, ,
\label{eq:inti3}
\eea
where the integral is taken over $SU(2)$ with the normalized Haar measure 
\bea
\int dg := \frac{1}{8\pi^2}\int_0^{2\pi} d\alpha \int_0^{2\pi} d\gamma 
\int_0^{\pi} d\beta \sin\beta \, .
\eea

Substituting the asymptote (\ref{eq:asymp}) for each matrix element 
$D^{J_i}_{M_iM_i'}(g)$ ($i=1,2,3$), the integral (\ref{eq:inti3}) becomes
\bea
\label{eq:3jint}
&&\int dg  \Big{(} \frac{1}{ 4 \pi J_1 \sqrt{\Delta_1}} \Big{)}^{1/2}
\Big{(} \frac{1}{ 4 \pi J_2 \sqrt{\Delta_2}} \Big{)}^{1/2}
\Big{(} \frac{1}{ 4 \pi J_3 \sqrt{\Delta_3}} \Big{)}^{1/2}
\crcr
&&\prod_{i=1}^3 \sum_{s_i=\pm1} e^{-\imath J_i (\alpha+\gamma)} \frac{1}{\sqrt{s_i \imath}}
 e^{\imath s_i\Big{(} \frac{\phi_i}{2}+J_i \bigl(\phi_i+ x_i \psi_i- y_i \omega_i) \bigr) 
\Big{)}}\;.
\eea
We expand (\ref{eq:3jint}), perform the integration over $\alpha$ and $\gamma$ and change variables 
from $\beta$ to $\xi$
\bea
 \frac{1}{2}\int_0^{\pi} \sin\beta d\beta = \frac{1}{2}
\int_{0}^{\pi} 2 \sin\frac{\beta}{2} \cos\frac{\beta}{2} d\beta = 
2 \int_0^1 \xi d\xi = \int_{0}^1 d(\xi^2)\;,
\eea
to rewrite it as 
\bea
 \delta_{\sum_i J_i x_i,0 }\; 
\delta_{\sum_i J_iy_i,0 } \;
 \Big{ [ } \int_0^1 d(\xi^2) \Big{]} 
 \Big{(}\frac{1} { (4\pi)^3 \,\prod_i J_i \sqrt{ \prod_i \Delta_i}} \Big{)}^{\frac{1}{2}}
 \sum_{s_i=\pm 1} \frac{1}{\sqrt{\prod_i s_i \imath^3 }}
e^{\imath \sum_i s_i \bigl( \frac{\phi_i}{2} + f_i  \bigr) } \; ,
\label{eq:startlast3}
\eea
where the index $i$ runs from $1$ to $3$, $\delta_{\sum_i J_ix_i ,0}$ is 
a  Kronecker symbols and $f_i$ is 
\bea
f_i &=& J_i \bigl[\phi_i + x_i \psi_i- y_i \omega_i \bigr] \; .
\eea 
We will derive the asymptotic behavior of eq. (\ref{eq:startlast3}) via an SPA 
with respect to $\xi^2$. Note that eq. (\ref{eq:inti3}) involves two distinct 
$3j$ symbols. If one attempts to first set $M'_i=M_i$, and obtain a representation of 
the square of a single $3j$ symbol, one encounters a very serious technical problem. We will see in 
the sequel that there are two saddle points $\xi^2_{\pm}$ contributing to the asymptotic behavior of
eq. (\ref{eq:startlast3}). If one starts by setting $M_i=M'_i$, one of the two saddle points  
$\xi^2_+=1$, and the second derivative in $\xi^2_+$ diverges. The contribution of this saddle point
does not evaluate by a simple Gaussian integration. 

The SPA evaluation of the general case, eq. (\ref{eq:startlast3}), is a very lengthy computation. 
We will perform it using the classical angular momentum vectors. For large representation 
index $J_i$ , there exists a classical angular momentum vector $\vec J_i$ in $\mathbb{R}^3$ 
of length $|\vec J_i|=J_i$ and projection 
on the $Oz$ axis (of unit vector $\vec n$) $\vec n\cdot \vec J_i=M_i$. A $3j$ symbol is then associated 
to three vectors, $\vec J_1,\vec J_2, \vec J_3$ with $|\vec J_i|=\vec J_i$ and 
$\vec n \cdot \vec J_i = M_i = x_iJ_i$. By the selection rules the quantum numbers $J_i$ respect the triangle
inequalities, and $M_1+M_2+M_3=0$. This translate into the condition that the vectors $\vec J_i$ form a 
triangle $\vec J_1+\vec J_2+\vec J_3=0$ (and $\vec n \cdot [\vec J_1+\vec J_2+\vec J_3]=0$). The asymptotic 
behavior of the $3j$ symbol writes in terms of the angular momentum vectors as

\begin{theorem}\label{th:3j}
 For large representation indices $J_i$ the $3j$ symbol has the asymptotic behavior
\bea
\left(\begin{array}{ccc}
 J_1&J_2&J_3\\
M_1&M_2&M_3
\end{array}\right) = \frac{1}{\sqrt{\pi (\vec n \cdot \vec S)}} \cos \Big{[}\sum_i 
\bigl(J_i+\frac{1}{2} \bigr)\Phi^i_{\vec n} 
+ (\vec n\cdot \vec J_1) \Psi^{13}_{\vec n}+(\vec n\cdot \vec J_2) \Psi^{23}_{\vec n}  +\frac{\pi}{4}\Big{]}\;,
\eea
with $\vec S = \vec J_1 \wedge \vec J_2= \vec J_2\wedge \vec J_3 = \vec J_3\wedge \vec J_1$, twice the area of the triangle
$\{\vec J_i\}$ and $\Phi^i_{\vec n}$ and $\Psi^{13}_{\vec n}$ and $\Psi^{23}_{\vec n}$ five 
angles defined as
\bea
&& \imath\Phi^i_{\vec n} = \ln \frac{ \vec n \cdot (\vec J_i \wedge \vec S) + \imath J_i ({\vec n} \cdot \vec S) }
{S \sqrt{(\vec n \wedge \vec J_i)^2 } } \; ,\crcr
&& \imath \Psi^{i3}_{\vec n}= \ln \frac{ (\vec n \wedge \vec J_i)\cdot (\vec n \wedge \vec J_3)+  
\imath \vec n \cdot (\vec J_3 \wedge \vec J_i)}{\sqrt{ (\vec n\wedge \vec J_i)^2(\vec n\wedge \vec J_3)^2 }  } \; ,
\qquad i=1,2 \; .
\eea
\end{theorem}

\begin{figure}[htb]
 \centerline{\includegraphics[width=2cm]{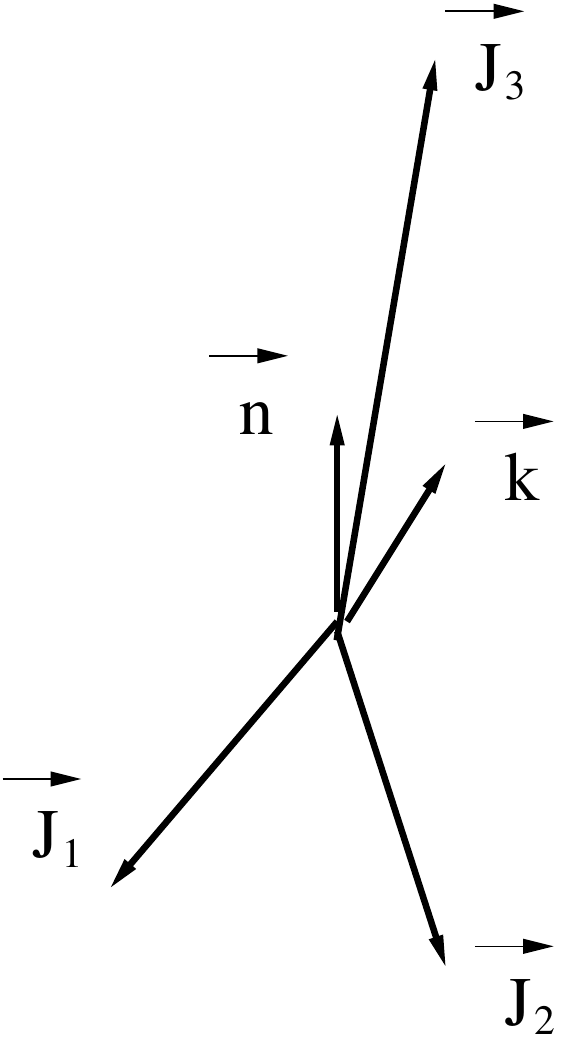}}
 \caption{Angular momentum vectors}
 \label{fig:mom}
\end{figure}

Before proceeding with the proof of Theorem \ref{th:3j} note that our starting equation (\ref{eq:inti3}) involves two
distinct $3j$ symbols. They are each associated to a triple of vectors, $\vec J_1,\vec J_2, \vec J_3$ 
($|\vec J_i|=\vec J_i$ and $\vec n \cdot \vec J_i = x_iJ_i$) and $\vec J_1',\vec J_2', \vec J_3'$ 
( $|\vec J'_i|= J_i$, $\vec n \cdot \vec J_i'=y_iJ_i$). Note that $|\vec J_i|=|\vec J'_i|$ hence 
the two triangles $\{\vec J_i\}$ and $\{\vec J'_i\}$ are congruent.
Consequently there exists a rotation which overlaps them. Under this rotation the normal vector $\vec n$ 
turns into the unit vector $\vec k$. All the geometrical information can therefore be encoded into an {\it unique} 
triple of vectors, henceforth denoted $\vec J_i$, and the {\it two} unit vectors $\vec n$ and 
$\vec k$ such that $|\vec J_i|=J_i$, $\vec n \cdot \vec J_i=x_iJ_i$ and $\vec k\cdot \vec J_i=y_iJ_i$ (see figure \ref{fig:mom}).

\medskip

\noindent{\bf Proof of Theorem \ref{th:3j}:} The proof follows the by now familiar routine of an SPA. We perform this
evaluation at fixed angular momenta, that is at fixed set of vectors $\vec J_i, \vec n, \vec k$.

\medskip

\noindent{\bf The dominant saddle points:} The saddle points governing the asymptotic
 behavior of eq. (\ref{eq:startlast3}) are solutions of the equation
\bea\label{eq:part1}
 0=\partial_{(\xi^2)} \sum s_i (\imath f_i) 
= \imath \sum_i s_i J_i [ \partial_{(\xi^2)} \phi_i + x_i \partial_{(\xi^2)} \psi_i-  y_i \partial_{(\xi^2)}\omega_i) ] \; .
\eea
A straightforward computation (see Appendix \ref{app:3jderiv}) yields 
\bea\label{eq:derivbun}
\partial_{(\xi^2)} \sum s_i (\imath f_i) = -\frac{\imath}{\xi^2(1-\xi^2)} \sum_i s_iJ_i \sqrt{\Delta_i} \; ,
\eea
hence the saddle point equation writes 
\bea\label{eq:3jsaddleready}
0 = s_1 J_1 \sqrt{\Delta_1}
+ s_2 J_2 \sqrt{\Delta_2}
+  s_3 J_3 \sqrt{\Delta_3}\;.
\eea
Introducing the angular momentum vectors the saddle point equation becomes after a short computation 
(see Appendix \ref{app:3jsaddle}) 
\bea\label{eq:3jsaddlefinal}
&&4 \xi^4 S^2 - 4\xi^2 \Big{\{ } S^2 + (\vec n\cdot \vec k) S^2  - (\vec n \cdot \vec S)(\vec k \cdot \vec S)\Big{\}}
\nonumber\\
&& + \Big{\{}
\bigl[1+(\vec n\cdot\vec k)\bigr]^2 S^2 - 2 (\vec n\cdot\vec S) (\vec k\cdot\vec S) 
\bigl[1+(\vec n\cdot\vec k)\bigr] \Big{\}}
 =0 \; ,
\eea
for all choices of signs $s_1, s_2$ and $s_3$.
Dividing by $4S^2$, eq. (\ref{eq:3jsaddlefinal}) factors as
\bea
 \Big{[} \xi^2 - \frac{1+(\vec n \cdot \vec k)}{2} \Big{]} 
\Big{[} \xi^2- \Big{(} \frac{1+(\vec n \cdot \vec k)}{2} - 
\frac{(\vec n \cdot \vec S)(\vec k \cdot \vec S)}{S^2} \Big{)} \Big{]}=0 \; ,
\eea
with roots 
\bea\label{eq:rootsxi}
 \xi^2_+ = \frac{1+(\vec n \cdot \vec k)}{2} \; , \qquad 
 \xi^2_-= \frac{1+(\vec n \cdot \vec k)}{2} - \frac{(\vec n \cdot \vec S)(\vec k \cdot \vec S)}{S^2} \; ,
\eea
again independent of the signs $s_1,s_2$ and $s_3$. To identify the terms contributing to the asymptotic 
of eq. (\ref{eq:startlast3}) for fixed $\vec J_i$, $\vec n$ and $\vec k$ one needs to evaluate 
$J_i \sqrt{\Delta_i}$ for each of the two roots $\xi^2_+$ and $\xi^2_-$. Using Appendix \ref{app:3jJDelta},
we have 
\bea
 J_i^2 \Delta_i^+  = \frac{1}{4} \Big{[} \vec J_i \cdot (\vec n \wedge \vec k) \Big{]}^2
\; , \qquad 
 J_i^2 \Delta_i^- = \frac{1}{4} \frac{\Big{\{}\vec J_i \cdot \Big{[} (\vec S \wedge \vec n)  
(\vec k \cdot \vec S) +  (\vec S \wedge \vec k)  (\vec n \cdot \vec S) \Big{]} \Big{\}}^2}{S^4} 
\; .
\eea
To any semiclassical state $\vec J_i$, $\vec n$, $\vec k$ we associate six signs, $\epsilon_i^+$ and $\epsilon_i^-$
defined by 
\bea\label{eq:signs}
J_i \sqrt{\Delta_i^+} = \epsilon_i^+ \frac{1}{2} \vec J_i\cdot (\vec n \wedge \vec k)\;, \qquad
J_i \sqrt{\Delta_i^-} = \epsilon_i^- \frac{1}{2} 
\frac{\vec J_i \cdot \Big{[} (\vec S \wedge \vec n)  (\vec k \cdot \vec S)
+  (\vec S \wedge \vec k)  (\vec n \cdot \vec S) \Big{]}}{S^2} \; .
\eea
Substituting $J_i \sqrt{\Delta_i^{\pm}}$ into the saddle point eq. (\ref{eq:3jsaddleready}) the latter becomes
\bea
\frac{1}{2} (\sum_i s_i \epsilon_i^{\pm} \vec J_i) \cdot \vec A^{\pm} \; ,
\eea
with $\vec A^+ = (\vec n \wedge \vec k)$ and $\vec A^{-}= \frac{\Big{[} (\vec S \wedge \vec n)  (\vec k \cdot \vec S)
+  (\vec S \wedge \vec k)  (\vec n \cdot \vec S) \Big{]} }{S^2}$. As, on the other hand, $\sum_i \vec J_i = 0$, 
we conclude that at fixed semiclassical state we have two saddle points $\xi^2_+$ and two saddle points $\xi^2_-$
contributing
\begin{itemize}
 \item The $\xi^2_+$ saddle point in the term $s_i = \epsilon_i^+$ and in the term $s_i = -\epsilon_i^+$
 \item The $\xi^2_-$ saddle point  in the term $s_i = \epsilon_i^-$ and in the term $s_i = -\epsilon_i^-$
\end{itemize}
The SPA evaluation of eq. (\ref{eq:startlast3}) is the sum of this four contributions.

\medskip

\noindent{\bf The second derivative:} The derivative of eq. (\ref{eq:derivbun}) with respect to $\xi^2$ yields
\bea
\partial_{(\xi^2)} [\partial_{(\xi^2)} \sum_i s_i  (\imath f_i)  ] &=& - \imath 
\partial_{(\xi^2)} \Big{(}\frac{1}{\xi^2(1-\xi^2)} \Big{)} 
\sum_i s_iJ_i \sqrt{\Delta_i} \crcr
&&-\frac{\imath}{\xi^2 (1-\xi^2)} \sum_i s_iJ_i
\frac{-(2\xi^2-1-x_i y_i)}{2\sqrt{\Delta_i}} \; ,
\eea
and the term in the first line cancels (due to the saddle point equation) when evaluating the second derivative at the critical points.
After Gaussian integration of the dominant saddle point contributions, the prefactor in the SPA approximation of
eq. (\ref{eq:startlast3}) writes
\bea
 \frac{1}{\sqrt{K}}\; , \qquad K= 32\;  \pi^2 s_1s_2s_3 \imath^3 J_1J_2J_3 \sqrt{\Delta_1 \Delta_2 \Delta_3 } 
\Bigl(- \partial^2_{(\xi^2)} \sum_i s_i (\imath  f_i ) \Bigr) \; .
\eea
The reminder of this paragraphs is devoted to the evaluation of the $K$ for the two roots $\xi^2_+$ and
$\xi^2_-$. Substituting the second derivative yields
\bea
K^{\pm}= - 16 \pi^2 s_1s_2s_3 \imath^4 \frac{J_1J_2J_3}{\xi^2_{\pm} (1-\xi^2_{\pm})} \sum_i s_i J_i 
\frac{\sqrt{\Delta_1\Delta_2\Delta_3}}{\sqrt{\Delta_i}}
(2\xi^2_{\pm}-1-x_i y_i )  \; .
\eea
Taking into account $s_1^2 s_2s_3= \epsilon^{\pm}_2\epsilon^{\pm}_3$, $K^{\pm}$  writes
\bea\label{eq:3jKpm}
K^{\pm }=  - (16 \pi^2) \frac{ \Big{[} \epsilon^{\pm}_2 \epsilon^{\pm}_3
J_2\sqrt{\Delta_2^{\pm}} J_3\sqrt{\Delta_3^{\pm}} 
\bigl[ (2 \xi_{\pm}^2 -1 )J_1^2 - J_1^{\vec n}  J_1^{\vec k} \bigr] 
+ \circlearrowleft_{123} \Big{]} }{\xi_{\pm}^2(1-\xi_{\pm}^2)} \; ,
\eea
where $\circlearrowleft_{123}$ denotes circular permutations on the indices $1$, $2$ and $3$.
Using eq. (\ref{eq:rootsxi}), the denominator evaluates, for the $\xi^2_+$ root, 
\bea
 \xi_+^2 (1- \xi_+^2) = \frac{1- (\vec n\cdot \vec k)^2}{4} \; ,
\eea
while the numerator computes to (see Appendix \ref{app:3jsecondderiv} 
for detailed computations and notations)
\bea
 \epsilon^{+}_2 \epsilon^{+}_3
J_2\sqrt{\Delta_2^{+}} J_3\sqrt{\Delta_3^{+}} 
\Bigl[ (2 \xi_{+}^2 -1 )J_1^2 - J_1^{\vec n}   J_1^{\vec k} \Bigr] 
+ \circlearrowleft_{123}   =  - \frac{1}{4} S^{\vec n} S^{\vec k} (\vec n \wedge \vec k)^2 \; ,
\eea
hence
\bea\label{eq:pres1}
 K^+ = 16 \pi^2 S^{\vec n} S^{\vec k} \; .
\eea

Evaluating the denominator in eq. (\ref{eq:3jKpm}) for $\xi^2_-$ yields 
\bea
&& \xi_-^2 (1- \xi_-^2) = \Big{(} \frac{1+(\vec n \cdot \vec k)}{2} -\frac{ S^{\vec n}  S^{\vec k}}{S^2}
 \Big{)} \Big{(}\frac{1-(\vec n \cdot \vec k)}{2} + \frac{ S^{\vec n}  S^{\vec k}}{S^2} \Big{)}\crcr
&& = \frac{1}{4} \Big{\{} (1-(\vec n\cdot \vec k)^2 + 4 (\vec n\cdot \vec k ) \frac{ S^{\vec n}  S^{\vec k}}{S^2}
-4 \frac{( S^{\vec n} S^{\vec k})^2}{S^4}
 \Big{\}} \; ,
\eea
while a lengthy computation (see Appendix \ref{app:3jsecondderiv}) shows that the numerator is
\bea
&& \epsilon^{-}_2 \epsilon^{-}_3
J_2\sqrt{\Delta_2^{-}} J_3\sqrt{\Delta_3^{-}} 
\Bigl[ (2 \xi_{-}^2 -1 )J_1^2 - J_1^{\vec n}   J_1^{\vec k} \Bigr] 
+ \circlearrowleft_{123}   = \crcr
&&= \frac{1}{4} S^{\vec n} S^{\vec k} 
\Big{\{} 1-(\vec n\cdot \vec k)^2 + 4 (\vec n\cdot \vec k ) \frac{ S^{\vec n}  S^{\vec k}}{S^2}
-4 \frac{( S^{\vec n} S^{\vec k})^2}{S^4}
 \Big{\}}
 \; ,
\eea
proving that  
\bea\label{eq:pres2}
 K^- = - 16 \pi^2 S^{\vec n} S^{\vec k} \; .
\eea

\medskip

\noindent{\bf Contribution of each saddle:} To evaluate the contribution of each saddle point to 
the asymptote of eq. (\ref{eq:startlast3}) we first evaluate 
\bea\label{eq:3jsaddleini}
\imath \sum_i s_i \bigl[\frac{\phi_i}{2}+ f_i \bigr] = \sum_i s_i \Big{[} \Bigl(J_i+\frac{1}{2} \Bigr) (\imath \phi_i^{\pm})
+x_i J_i (\imath \psi_i^{\pm}) -y_i J_i (\imath \omega_i^{\pm}) \Big{]} \; .
\eea
Recall that for a fixed semiclassical state only the terms with $s_i$ equal to $\epsilon_i^{+}$, $-\epsilon_i^{+}$,
$\epsilon_i^{-}$ and $-\epsilon_i^{-}$ contribute. We substitute $x_3J_3= -x_2J_2-x_1J_1$  and $y_3J_3 = -y_1J_1 - y_2J_2$ into 
eq. (\ref{eq:3jsaddleini}) to bring it into the form
\bea\label{eq:3jsad}
\pm &&\Big{\{}\sum_i \Bigl(J_i+\frac{1}{2} \Bigr) (\imath \epsilon_i^{\pm} \phi_i^{\pm} ) + x_1J_1 
(\imath \epsilon_1^{\pm} \psi_1^{\pm} - \imath \epsilon_3^{\pm} \psi_3^{\pm}) + 
x_2J_2 (\imath \epsilon_2^{\pm} \psi_2^{\pm} - \imath \epsilon_3^{\pm} \psi_3^{\pm}) \crcr
&&-y_1J_1 
(\imath \epsilon_1^{\pm} \omega_1^{\pm} - \imath \epsilon_3^{\pm} \omega_3^{\pm}) -
y_2J_2 (\imath \epsilon_2^{\pm} \omega_2^{\pm} - \imath \omega_3^{\pm} \psi_3^{\pm})
\Big{\}} \; ,
\eea
where $\phi_i^{\pm}$, $\psi_i^{\pm}$ and $\omega_i^{\pm}$ are the angles $\phi_i$, $\psi_i$ and $\omega_i$ evaluated at
$\xi^2_+$ and $\xi^2_-$. For each choice $+$ or $-$ in the accolades, one must count {\it both} 
choices of the overall sign. The angles $\phi_i^{\pm}$, $\epsilon_1^{\pm} \psi_1^{\pm} - \epsilon_3^{\pm} \psi_3^{\pm}$,
etc. are evaluated by a rather involved computation in Appendix \ref{app:3jangles}. The end results are synthesized below
\bea\label{eq:amg}
&& \imath \epsilon_i^{\pm} \phi_i^{\pm} = \imath \Phi^i_{\vec n} \mp \imath \Phi^i_{\vec k} \; , \qquad
\imath\Phi^i_{\vec n} = \ln \frac{ \vec n \cdot (\vec J_i \wedge \vec S) + \imath J_i S^{\vec n} }
{S \sqrt{(\vec n \wedge \vec J_i)^2 } }
\crcr
&&  \imath  \epsilon_j^{\pm} \psi_j^{\pm}- \imath  \epsilon_3^{\pm} \psi_3^{\pm}  =  \imath \Psi^{j3}_{\vec n}
\; , \qquad
 \imath \Psi^{j3}_{\vec n} = \ln \frac{ (\vec n \wedge \vec J_j)\cdot (\vec n \wedge \vec J_3) 
 +  \imath \vec n \cdot (\vec J_3 \wedge \vec J_j)}{\sqrt{ (\vec n\wedge \vec J_j)^2(\vec n\wedge \vec J_3)^2 }  }
\; , \quad j=1,2 \; ,\crcr
&&\imath  \epsilon_j^{\pm} \omega_j^{\pm}- \imath  \epsilon_3^{\pm} \omega_3^{\pm}  =  \pm \imath \Psi^{j3}_{\vec k}
\; .
\eea

Substituting eq. (\ref{eq:amg}) into eq. (\ref{eq:3jsad}) yields
\bea\label{eq:pres3}
 \pm && \Big{\{} \sum_i  \bigl(J_i+\frac{1}{2} \bigr) \bigl(\imath \Phi^i_{\vec n} \mp \imath \Phi^i_{\vec k}\bigr) 
+ (\vec n \cdot \vec J_1)\imath \Psi^{13}_{\vec n} + (\vec n \cdot \vec J_2)\imath \Psi^{23}_{\vec n}
\crcr
&& \mp (\vec k \cdot \vec J_1)\imath \Psi^{13}_{\vec k} \mp (\vec k \cdot \vec J_2)\imath \Psi^{23}_{\vec k} 
= \pm (\Omega_{\vec n} \mp \Omega_{\vec k}) 
\; ,
\eea
where $\Omega_{\vec n}$ denotes 
\bea
 \imath \Omega_{\vec n}= \sum_i  \bigl(J_i+\frac{1}{2} \bigr) \imath \Phi^i_{\vec n} 
+ (\vec n \cdot \vec J_1)\imath \Psi^{13}_{\vec n} + (\vec n \cdot \vec J_2)\imath \Psi^{23}_{\vec n}\; .
\eea

\medskip

\noindent{\bf Final evaluation:} We put together eq. (\ref{eq:pres1}), (\ref{eq:pres2}) 
and (\ref{eq:pres3}) and, noting that the two contributions form the saddle $\xi^2_-$ are 
complex conjugate to one another we obtain
\bea
&&\left(\begin{array}{ccc}
 J_1&J_2&J_3\\
M_1&M_2&M_3
\end{array}\right)\left(\begin{array}{ccc}
 J_1&J_2&J_3\\
M'_1&M'_2&M'_3
\end{array}\right) \crcr
&&\approx \frac{1}{\sqrt{\pi (\vec n\cdot \vec S) }}
\frac{1}{\sqrt{\pi (\vec k\cdot \vec S) }} \frac{1}{4}
\Big{(} e^{\imath (\Omega_{\vec n}- \Omega_{\vec k}) } + e^{-\imath (\Omega_{\vec n}- \Omega_{\vec k}) }
+\imath e^{\imath (\Omega_{\vec n} + \Omega_{\vec k}) }
-\imath e^{-\imath (\Omega_{\vec n} + \Omega_{\vec k}) }
\Big{)} \; .
\eea
Taking into account
\bea
\frac{1}{4}
\Big{(} e^{\imath (\Omega_{\vec n}- \Omega_{\vec k}) } + e^{-\imath (\Omega_{\vec n}- \Omega_{\vec k}) }
+\imath e^{\imath (\Omega_{\vec n} + \Omega_{\vec k}) }
-\imath e^{-\imath (\Omega_{\vec n} + \Omega_{\vec k}) }
\Big{)} = \cos \Bigl(\Omega_{\vec n} + \frac{\pi}{4} \Bigr) \cos \Bigl(\Omega_{\vec k} + \frac{\pi}{4}\Bigr)
\;,
\eea
Theorem \ref{th:3j} follows.

\qed

\section{Conclusion}
\label{concl}

Using the EM SPA method we have determined the asymptotic behaviors at large spin $J$ 
of Wigner matrix elements, Wigner $3j$ symbols and the character $\chi^J(g)$ of an $SU(2)$ 
group element $g$.

By far the most surprising fact about this computation is that our formula for the character $\chi^J(g)$
is exact. SPA reproducing the exact result for integrals are usually the consequence of a
Duistermaat Heckman \cite{DH1,DH2}localization property (one of the most famous example of this being the 
Harish Chandra Itzykson Zuber integral \cite{itzub}). Recall that the Duistermaat-Heckman
theorem states that a phase space integral
\bea\label{eq:frfrt}
 \int \Omega \; e^{-\imath H(p,q)}\; ,
\eea
where $\Omega$ is the Liouville form, equals its leading order SPA estimation if the flow of the Hamiltonian 
vector field ${\bf X}$ ($i_{{\bf X} } \Omega=dH$) is $U(1)$. To our knowledge all integrals
exhibiting a localization property (i.e. equaling their leading order SPA approximation) fall in 
(some generalization of) this case. Note that the character of an $SU(2)$ group element can be expressed 
directly as a double integral by 
\bea\label{eq:chichi}
\chi^J(g) &=& \sum_{M,t} e^{h(J,M,t)} \approx \frac{J}{2\pi} \int du dx \; \sqrt{K(x,x,u)} e^{J f(x,x,u)} 
+ \text{E.M.}+ \text{S.} \; ,
\eea
where E.M. denotes corrections coming from the Euler Maclaurin approximation, and $S$ the corrections coming from 
sub leading terms in the Stirling approximation. The double integral in equation (\ref{eq:chichi}) is of the 
correct form, with symplectic form $\Omega = \sqrt{K(x,x,u)} dx \wedge du$ and Hamiltonian $f(x,x,u)$ 
generating the Hamiltonian flow
\bea
\frac{du}{d\rho} &=& 
\sqrt{\frac{u^2 (1+x-u)(1-x-u)}{1-x^2}}\,  \ln\left\{  
e^{-\imath(\alpha + \gamma)} \frac{(1+x)(1-x-u)}{(1-x)(1+x-u)}  
\right\} \\
\frac{dx}{d\rho} &=& -\sqrt{\frac{u^2 (1+x-u)(1-x-u)}{1-x^2}} \,  \ln\left\{  
e^{\imath \pi} \frac{(1-\xi^2)}{\xi^2}\frac{(1-x-u)(1+x-u)}{u^2}  
\right\} \; .
\eea
Our result can be explained if first, the above flow is $U(1)$ (thus the SPA 
of the double integral is exact) and second the EM and Stirling correction terms cancel, 
$\text{E.M.}+ \text{S.}=0$. The alternative, namely that the flow is not $U(1)$ would
require an even more subtle cancellation of the sub leading correction terms. Either way,
the exact result for the character we derive in this paper deserves further investigation.

\section*{Acknowledgements}

Research at Perimeter Institute is supported by the Government of Canada
through Industry Canada and by the Province of Ontario through the Ministry
of Research and Innovation.

\section*{Appendix} 

\appendix

\renewcommand{\theequation}{\Alph{section}.\arabic{equation}}
\setcounter{equation}{0}

In these appendices we detail various technical points and computations.

\section{The Stirling approximation}
\label{app:stir}

We detail here the passage from eq. (\ref{eq:DfromF}) to eq. (\ref{eq:dlam}).
Our starting point is 
\bea\label{eq:DfromF1}
D^{J}_{MM'}(\alpha,\beta,\gamma) \approx \int dt\; F(J,M,M',t) \; ,
\eea
with 
\bea\label{eq:Fmare}
&&F(J,M,M',t) = e^{\imath\pi t} e^{-\imath\alpha M} e^{-\imath \gamma M'} \xi^{2J+M-M'-2t} \eta^{2t-M+M'} \times
\nonumber\\
&&\frac{\sqrt{\Gamma(J+M+1)\Gamma(J-M+1)\Gamma(J+M'+1)\Gamma(J-M'+1)}}{ \Gamma(J+M-t+1)
\Gamma(J-M'-t+1)\Gamma(t+1)\Gamma(t-M+M'+1)} \; .
\eea
We use the Stirling formula
\bea
\Gamma(n+1)= n! \approx \sqrt{2\pi n} \left(\frac{n}{e}\right)^{n} = \sqrt{2\pi n }\, e^{ n\ln n - n} \; ,
\eea
for all $\Gamma$ functions and rescaled variables $M=xJ$, $M'=yJ$, $t=uJ$. Collecting all prefactors yields
\bea
\Big{(} \frac{\sqrt{ (2\pi)^{4} J^4 (1+x)(1-x)(1+y)(1-y) }}
{ (2\pi)^4 J^4 (1+x-u)(1-y-u) (u) (u-x+y)} \Big{)}^{\frac{1}{2}} = \frac{1}{2\pi J} \sqrt{K(x,y,u)} \; ,
\eea
and $K(x,y,u)$ takes the form in eq. (\ref{eq:primK}). The ``-n'' terms in the Stirling approximation 
add to
\bea
&& \frac{1}{2} \Big{\{} - J(1+x) - J(1-x) - J(1+y) - J(1-y)
\Big{\}} \nonumber\\
&& - \Big{\{} - J(1+x-u) - J(1-y-u) - Ju - J(u-x+y)
\Big{\}} =0 \; ,
\eea
which also implies that the coefficient of $\ln J$ in the exponent cancels. 
The contribution of the $\Gamma$ functions eq. (\ref{eq:Fmare}) is therefore 
\bea\label{eq:stirstir}
&& \frac{J}{2} \Big{\{} (1+x)  \ln (1+x)  +  (1-x) \ln (1-x)+  (1+y) \ln (1+y)  +  (1-y) \ln (1-y) 
\Big{\}} \nonumber\\
&& - J \Big{\{}  (1+x-u) \ln (1+x-u) + (1-y-u) \ln (1-y-u)  \\
&&+  u \ln ( u ) +  (u-x+y ) \ln (u-x+y) 
\Big{\}} \; .\nonumber
\eea
The substitution of eq. (\ref{eq:stirstir}) into eq. (\ref{eq:Fmare}) yields 
\begin{equation}
F(J,xJ,yJ,uJ)
\approx  \frac{1}{2\pi J} \sqrt{K(x,y,u)} e^{J f(x,y,u)} \; ,
\end{equation}
where $f(x,y,u)$ takes the form in eq. (\ref{eq:primf}), and
\bea
D^{J}_{MM'}(\alpha,\beta,\gamma) \approx \int dt\; F(J,M,M',t) \approx \int dt \frac{1}{2\pi J} 
\sqrt{K(x,y,u)} e^{J f(x,y,u)} \; ,
\eea
which reproduces eq. (\ref{eq:dlam}) after changing the integration variable to $u = \frac{t}{J}$.

\section{Evaluations on the critical set.}
\label{app:eval}

In this appendix we present the various evaluations relevant for the 
proof of Theorem \ref{theo1}. We start by some preliminary computations. Recall that
\bea
 \Delta = (1-\xi^2)(\xi^2-xy)-\frac{(x-y)^2}{4} \ge 0\; .
\eea
As a preliminary we compute the absolute values of the four complex numbers  
\bea\label{eq:cruc}
&& u_{\pm} = 1- \xi^2 +\frac{x-y}{2} \pm \imath \sqrt{\Delta} \; , \qquad
u_{\pm}-x+y = 1-\xi^2 - \frac{x-y}{2} \pm   \imath \sqrt{\Delta } \; , \nonumber\\
&& 1+x-u_{\pm} =  \xi^2 + \frac{x+y}{2} \mp  \imath \sqrt{ \Delta  } \; , \qquad
1-y-u_{\pm} = \xi^2 - \frac{x+y}{2} \mp  \imath \sqrt{  \Delta   } \; ,
\eea
which are  
\bea\label{eq:abs}
&& |u_{\pm}|^2 =   (1-\xi^2) (1 + x)(1 - y)  \; , \qquad
|u_{\pm}-x+y|^2 = (1-\xi^2) (1-x)(1+y)  \; , \nonumber\\
&& |1+x-u_{\pm}|^2 =  \xi^2 (1+x)(1+y)  \; , \qquad
|1-y-u_{\pm} |^2 =  \xi^2 (1-x)(1-y) \; .
\eea

\subsection{Evaluation of $f$ at the critical points} 
\label{app:critf}
To establish eq. (\ref{eq:ffin}) and (\ref{eq:phichiom}), 
we evaluate eq. (\ref{eq:reducedf}) at $u_{\pm}$
\bea
&&
f(x,y,u_{\pm}) = -\imath\alpha x -\imath \gamma y + (2+x-y) \ln \xi
+ (-x+y) \ln \eta \nonumber\\
&&+ \frac{1}{2}(1-x) \ln(1-x) +  \frac{1}{2}(1+x) \ln(1+x)
 +  \frac{1}{2}(1-y) \ln(1-y) + \frac{1}{2}(1+y) \ln(1+y) \nonumber\\
&&-  (1+x)\ln(1+x-u_{\pm}) - (1-y) \ln(1-y-u_{\pm})  - (-x+y) \ln(u_{\pm}-x+y) \; .
\eea
The real part of $f(x,y,u_{\pm})$ is
\bea
&&\Re f(x,y,u_{\pm}) = (2+x-y) \ln \xi
+ (-x+y) \ln \eta \nonumber\\
&&+ \frac{1}{2}(1-x) \ln(1-x) +  \frac{1}{2}(1+x) \ln(1+x)
 +  \frac{1}{2}(1-y) \ln(1-y) + \frac{1}{2}(1+y) \ln(1+y) \nonumber\\
&&-  (1+x)\ln|1+x-u_{\pm}| - (1-y) \ln|1-y-u_{\pm}|  - (-x+y) \ln|u_{\pm}-x+y| \; ,
\eea
and substituting the absolute values computed in eq. (\ref{eq:abs}) yields
\bea
&&\Re f (x,y,u_\pm) = (2+x-y) \ln \xi + (-x+y) \ln \eta \crcr
&&+ \frac{1}{2}(1-x) \ln(1-x) +  \frac{1}{2}(1+x) \ln(1+x)
+  \frac{1}{2}(1-y) \ln(1-y) + \frac{1}{2}(1+y) \ln(1+y) \crcr
&&-  \frac{(1+x)}{2}\ln\Big{[} \xi^2 (1+x)(1+y) \Big{]} - \frac{(1-y)}{2} \ln\Big{[}  \xi^2 (1-x)(1-y)\Big{]} \crcr
&&- \frac{(-x+y)}{2} \ln\Big{[} (1-\xi^2) (1-x)(1+y)  \Big{]} \; .
\eea
Recalling that $1-\xi^2 = \eta^2$ we note that the coefficients of both $\ln \xi$ and $\ln (1-\xi^2)$ cancel. Furthermore, a
direct inspection shows that the coefficients of all $\ln(1-x)$, $\ln(1+x)$, 
$\ln(1-y)$ and $\ln(1+y)$ cancel. Hence 
\bea
 \Re f (x,y,u_\pm) = 0 \; .
\eea
Therefore $f(x,y,u_{\pm})$ is a purely imaginary number
\bea
f(x,y,u_{\pm})& =& -\imath\alpha x -\imath \gamma y 
-  (1+x)\ln\frac{1+x-u_{\pm}}{|1+x-u_{\pm}|} - 
(1-y) \ln\frac{1-y-u_{\pm}}{|1-y-u_{\pm}|}  \nonumber\\
&&- (-x+y) \ln\frac{u_{\pm}-x+y}{|u_{\pm}-x+y|} \; .
\eea
which assumes the form
\bea
 f(x,y,u_{\pm}) = -\imath\alpha x -\imath \gamma y \pm \imath \Big{(} \phi + x\psi - y\omega \Big{)} \; ,
\eea
where the three angles $\phi$, $\psi$ and $\omega$ read 
\bea\label{eq:angint}
&&  \imath \phi =  - \ln\frac{(1+x-u_+)}{|1+x-u_+|}- \ln \frac{(1-y-u_+)}{|1-y-u_+|} \; , \nonumber \\
&&  \imath \psi =  - \ln\frac{1+x-u_+}{|1+x-u_+|}  + \ln \frac{(u_+ -x+y)}{| u_+ -x+y |}  \; , \nonumber\\
&& \imath \omega=  -  \ln \frac{( 1-y-u_+ ) }{ |1-y-u_+| } +  \ln \frac{u_+ -x+y}{|u_+ -x+y|} \; .
\eea
As the two roots $u_+$ and $u_-$ are complex conjugate, one can absorb the various 
signs in eq. (\ref{eq:angint}) and write
\bea
&&\imath \phi = \ln \frac{ ( 1+x-u_-) (1-y-u_- )}{  |1+x-u_-| |1-y-u_-|} \; , 
\quad \imath \psi =  \ln\frac{ (1+x-u_- ) (u_+ -x+y) }{|1+x-u_-| | u_+ -x+y |}  \; , \nonumber\\
&& \imath \omega=  \ln \frac{( 1-y-u_- ) (u_+ -x+y) }{ |1-y-u_-| |u_+ -x+y|} \; .
\eea
One by one $\phi$, $\psi$ and $\omega$ compute by substituting eq. (\ref{eq:cruc}) 
and eq. (\ref{eq:abs}) to
\bea
&& \imath \phi = \ln \frac{ \bigl( \xi^2 + \frac{x+y}{2} +  \imath \sqrt{ \Delta  }  \bigr) 
\bigl( \xi^2 - \frac{x+y}{2} +  \imath \sqrt{  \Delta   }\bigr) }
{ \sqrt{\xi^4 (1-x^2) (1-y^2 )} }  \nonumber\\
&& = \ln \frac{ \xi^4 -\frac{(x+y)^2}{4} + 2\xi^2 \imath \sqrt{\Delta} -(1-\xi^2)(\xi^2-xy) + \frac{(x-y)^2}{4} }
{ \sqrt{\xi^4 (1-x^2) (1-y^2 )} }\nonumber\\
&&= \ln \frac{2\xi^2 -1 - xy + 2\imath \sqrt{\Delta}}{\sqrt{(1-x^2) (1-y^2 )}} \; ,
\eea
and
\bea
&& \imath \psi  = 
\ln \frac{ \bigl( \xi^2 + \frac{x+y}{2} +  \imath \sqrt{ \Delta  }  \bigr) 
\bigl( 1-\xi^2 - \frac{x-y}{2} +   \imath \sqrt{\Delta } \bigr)    } 
{\sqrt{\xi^2 (1-\xi^2)  (1-x^2) (1+y)^2 } } \nonumber\\
&&=\ln \frac{\xi^2(1-\xi^2) + \frac{x+y}{2} -x\xi^2 -\frac{x^2-y^2}{4} -(1-\xi^2)(\xi^2-xy) + 
\frac{(x-y)^2}{4} +\imath (1+y) \sqrt{\Delta}  } 
{\sqrt{\xi^2 (1-\xi^2)  (1-x^2) (1+y)^2 }} \nonumber\\
&&=\ln \frac{ -x(1+y) \xi^2  + \frac{x+y}{2}+ xy + \frac{y^2-xy}{2} +\imath (1+y) \sqrt{\Delta}  } 
{\sqrt{\xi^2 (1-\xi^2)  (1-x^2) (1+y)^2 }}  \nonumber\\
&&=\ln \frac{ \frac{x+y}{2} -x \xi^2 +\imath \sqrt{\Delta}  } 
{\sqrt{\xi^2 (1-\xi^2)  (1-x^2) }} \; ,
\eea
and finally
\bea
&& \imath \omega=  \ln \frac{
\bigl( \xi^2 - \frac{x+y}{2} +  \imath \sqrt{  \Delta   }\bigr) 
\bigl( 1-\xi^2 - \frac{x-y}{2} +   \imath \sqrt{\Delta } \bigr) }{ \sqrt{ \xi^2 (1-\xi^2)  (1-y^2) (1-x)^2 } }  
\nonumber\\
&& = \ln 
\frac{\xi^2(1-\xi^2) -\frac{x+y}{2} +y\xi^2 + \frac{x^2-y^2}{4} -(1-\xi^2)(\xi^2-xy) + 
\frac{(x-y)^2}{4} +\imath(1-x)\sqrt{\Delta}   }
{\sqrt{ \xi^2 (1-\xi^2)  (1-y^2) (1-x)^2 }} \nonumber\\
&&= \ln \frac{y(1-x)\xi^2  -\frac{x+y}{2} + xy  +\frac{x^2-xy}{2}  + \imath(1-x)\sqrt{\Delta}  }
{\sqrt{ \xi^2 (1-\xi^2)  (1-y^2) (1-x)^2 }} \nonumber\\
&&= \ln \frac{-\frac{x+y}{2}+y\xi + \imath \sqrt{\Delta} }{\sqrt{ \xi^2 (1-\xi^2)  (1-y^2) } } \; .
\eea

\subsection{Evaluation of the second derivative.}
\label{app:critderiv}
From eq. (\ref{eq:uhess}) we have
\bea
\label{eq:uhessapp}
 -\partial_u^2 f(x,y,u) = \frac{1}{1+x-u} +\frac{1}{1-y-u} +\frac{1}{u} +\frac{1}{u-x+y} \; .
\eea
Each term evaluates at the critical points as
\bea\label{eq:fractt}
&& \frac{1}{1+x-u_\pm } = \frac{1+x-u_{\mp}}{|1+x-u_{\pm}|^2} 
= \frac{\xi^2 + \frac{x+y}{2}\pm \imath \sqrt{\Delta}}{ \xi^2(1+x)(1+y)} \crcr
&& \frac{1}{1-y+u_\pm } = \frac{1-y+u_{\mp}} {|1-y+u_{\mp}|^2} 
=\frac{\xi^2 - \frac{x+y}{2}\pm \imath \sqrt{\Delta}}{ \xi^2(1-x)(1-y)} \crcr
&& \frac{1}{u_\pm -x+y} =\frac{u_\mp -x+y}{|u_\mp -x+y|^2}
= \frac{(1-\xi^2)- \frac{x-y}{2}\mp \imath \sqrt{\Delta} }{(1-\xi^2)(1-x)(1+y)   } \crcr
&& \frac{1}{u_\pm} =\frac{u_{\mp}}{|u_{\mp}|^2}
= \frac{(1-\xi^2) + \frac{x-y}{2}\mp \imath \sqrt{\Delta} }{(1-\xi^2)(1+x)(1-y)   } \; .
\eea
The real part of (\ref{eq:uhessapp}) is therefore
\bea
&& \frac{\xi^2 + \frac{x+y}{2}}{ \xi^2(1+x)(1+y)} +
 \frac{\xi^2 - \frac{x+y}{2}}{ \xi^2(1-x)(1-y)} + \nonumber\\
&& \frac{(1-\xi^2)- \frac{x-y}{2}}{(1-\xi^2)(1-x)(1+y)   }+
\frac{(1-\xi^2) + \frac{x-y}{2}}{(1-\xi^2)(1+x)(1-y)   } \; ,
\eea
and computes further to 
\bea
\Re (-\partial^2_u f)|_{x,y,u_{\pm}} = \frac{4\Delta}{(1-x^2)(1-y^2)\xi^2(1-\xi^2)}.
\eea
The imaginary part of eq. (\ref{eq:uhessapp}) is 
\bea
 \pm \imath \sqrt{\Delta}&& \Big{(}\frac{1}{ \xi^2(1+x)(1+y)} + \frac{1}{ \xi^2(1-x)(1-y)} \crcr
&&  - \frac{1}{(1-\xi^2)(1-x)(1+y)   } -\frac{1}{(1-\xi^2)(1+x)(1-y)  }\Big{)} \; ,
\eea
which finally computes to 
\bea
\Im (-\partial^2_u f)|_{x,y,u_{\pm}} = 
 \pm  \imath 2 \sqrt{\Delta} \; \frac{ 1- 2 \xi^2 -  xy }{(1-x^2)(1-y^2) \xi^2 (1-\xi^2)} \; .
\eea

\subsection{Evaluation of $K$}
\label{app:critK}
The prefactor $K|_{x,y,u_{\pm}}$ is
\bea
  K = \frac{ \sqrt{(1-x^2)(1-y^2) } }
 {(1+x-u_{\pm})(1-y-u_{\pm})(u_{\pm})(u_{\pm}-x+y) } \; ,
\eea
which is, using eq. (\ref{eq:fractt}),
\bea
K=&& \frac{\sqrt{(1-x^2)(1-y^2) } }{\xi^4 (1-\xi^2)^2 (1-x^2)^2(1-y^2)^2} \crcr
&& \Big{(}\xi^2 + \frac{x+y}{2}\pm \imath \sqrt{\Delta} \Big{)}
 \Big{(} \xi^2 - \frac{x+y}{2}\pm \imath \sqrt{\Delta} \Big{)} \crcr
&&\Big{(} (1-\xi^2)- \frac{x-y}{2}\mp \imath \sqrt{\Delta} \Big{)}
 \Big{(} (1-\xi^2) + \frac{x-y}{2}\mp \imath \sqrt{\Delta}\Big{)} \; ,
\eea
and a straightforward computation proves eq. (\ref{eq:kfac}).

\section{Real saddle points}
\label{app:delta<0}

\begin{figure}[htb] 
 \centering
\begin{minipage}[t]{0.8\textwidth}
\centering
\includegraphics[angle=0, width=4.2cm, height=4.2cm]{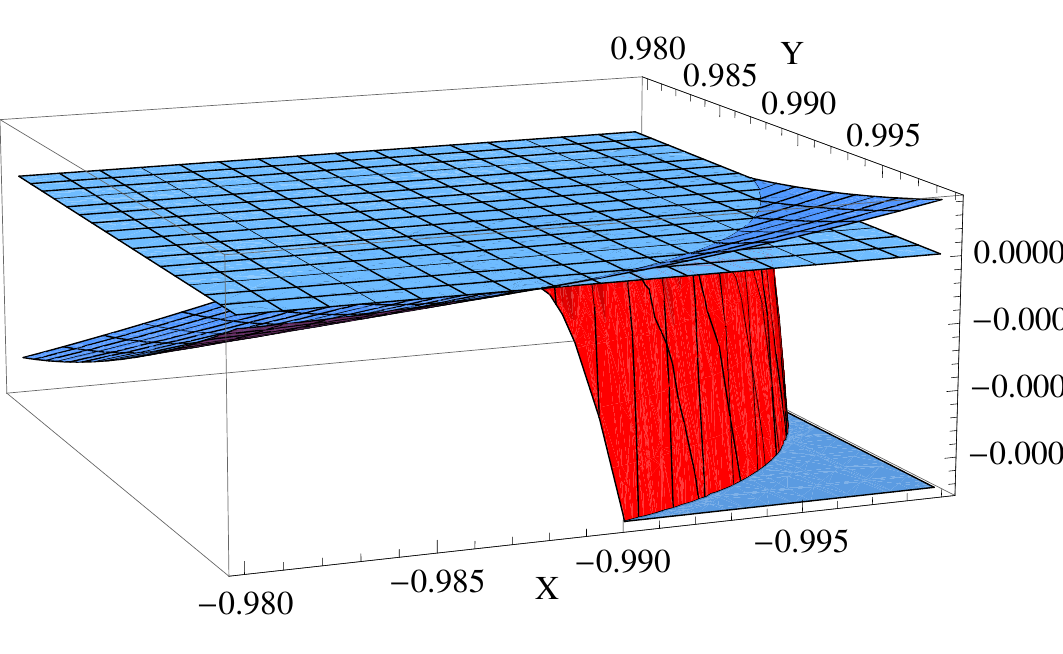}\hspace{0.1cm} 
\includegraphics[angle=0, width=4.2cm, height=4.2cm]{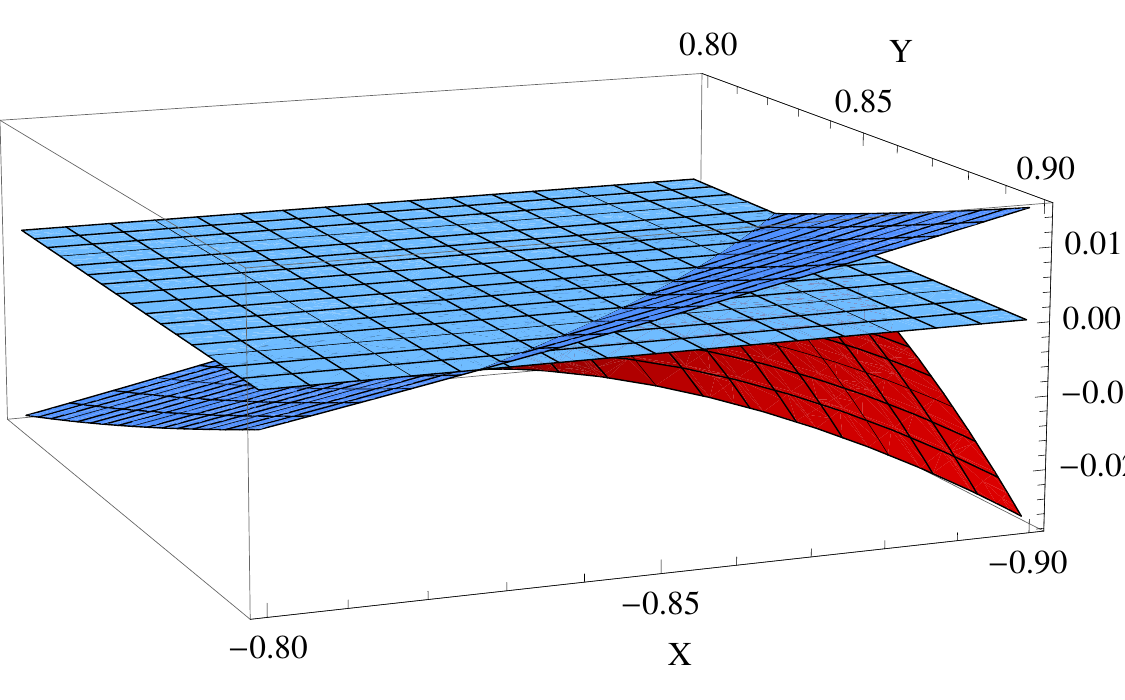}\hspace{0.2cm}
\includegraphics[angle=0, width=4.2cm, height=4.2cm]{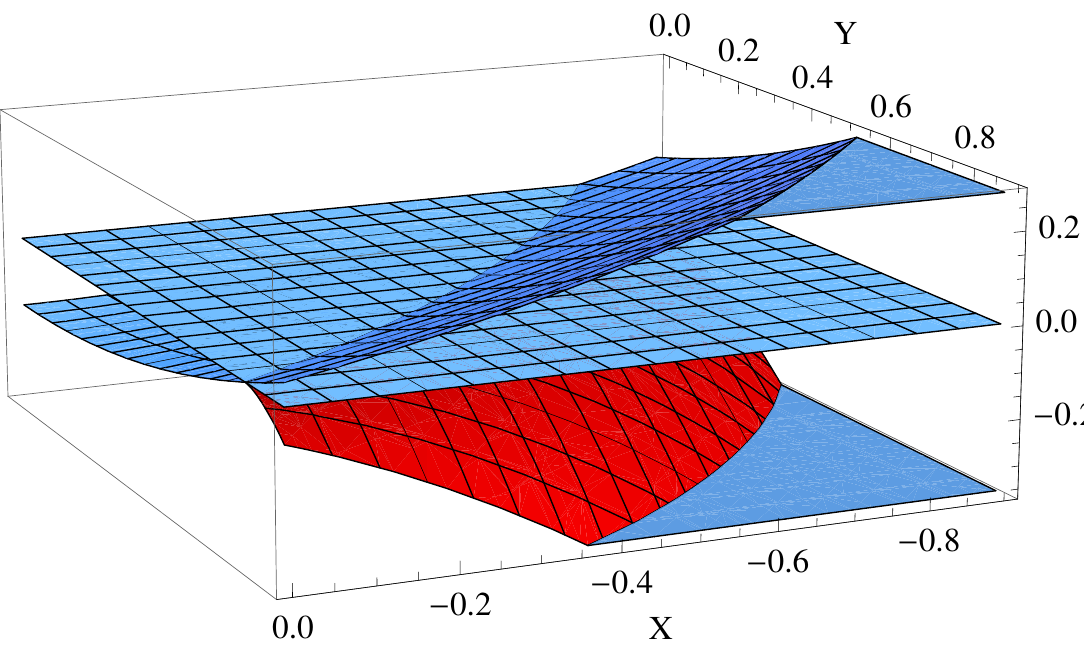}
\put(-80,24){$\Xi$}
\put(-170,37){$\Xi$}
\put(-320,34){$\Xi$}
\caption{ \small The function $\Xi= \Phi  + x \Psi -y \Omega$ (red)  is negative,
and vanishes (plane $z=0$, light blue) when $\Delta'$ (dark blue) vanishes, 
for $\xi=0.1,0.5$ and $0.9$, from the left to the right.}
\label{fig}
\end{minipage}
\end{figure}

In this section we present the SPA evaluation of a matrix element with
\bea
 \Delta = (1-\xi^2)(\xi^2-xy) - \frac{(x-y)^2}{4} <0 \; .
\eea
For convenience we denote $\Delta'=-\Delta>0$.
In this range of parameters the two saddle points
\bea
u_{\pm}= h_{\pm}(x;y) = (1-\xi^2) + \frac{x-y}{2} \pm  \sqrt{\Delta' } \; ,
\eea
are real. For simplicity suppose that $0<x\le y<1$. A straightforward computation shows that $0<u_-<u_{+}<1-y$, hence both roots
are in the integration interval. Using the results of Appendix \ref{app:critf}, the function evaluates at the two saddle points as
\bea
f|_{u_{\pm}} = -\imath\alpha x -\imath \beta y \pm (\Phi + x\Psi-y\Omega) \; ,
\eea
with 
 \bea
&&\Phi =\ln \frac{(2\xi^2 -1- xy + 2\sqrt{\Delta'})}{\sqrt{(1-x^2)(1-y^2)}},\\
&&\Psi = \ln \frac{(-x \xi^2 +\frac{x+y}{2} +  \sqrt{\Delta'})}{\sqrt{\xi^2(1-\xi^2)(1-x^2)}},\\
&& \Omega = \ln \frac{(\xi^2 y -\frac{x+y}{2} + \sqrt{\Delta'})}{\sqrt{\xi^2(1-\xi^2)(1-y^2)}}.
\eea
From Appendix \ref{app:critderiv}, we obtain
\bea\label{eq:realderiv}
 - \partial^2_u f \Big{|}_{u_\pm} = \frac{-4\Delta' \mp 2\sqrt{\Delta'}(2\xi^2-1-xy)}{\xi^2(1-\xi^2)(1-x^2)(1-y^2)}
\; ,
\eea
which shows in particular that the maximum of $f$ is $u_-$ (as $ - \partial^2_u f|_{u_-} <0$), and the SPA is 
dominated by the latter. In Figure (\ref{fig}) below we represent the function $\Xi = \Phi  + x \Psi -y \Omega$ 
as a function of $x$ and $y$, 

The prefactor writes,  using Appendix \ref{app:critK},
\bea
K|_{u-} = \frac{-\sqrt{(1-x^2)(1-y^2) }\Big{(} 2 \xi^2-1-xy - 2 \sqrt \Delta\Big{)}^2}{\xi^2 (1-\xi^2) (1-x^2)^2(1-y^2)^2}
\; ,
\eea
hence we get the asymptotic estimate
\bea\label{eq:delt<0est}
D^J_{x J,y J} (\alpha,\beta,\gamma)
&\approx& -
\frac{1}{\sqrt{2\pi J}} \,
\Big{(} \frac{1}{2 \sqrt{\Delta'}} \Big{)}^{1/2}  
e^{-\imath\alpha J x -\imath\gamma Jy}
e^{-\frac{\Phi}{2}}
e^{ - J \bigl( \Phi + x\Psi-y\Omega \bigr) } \; ,
\eea
which indeed is suppressed for large $J$.

The case $\Delta'=0$ is special.  A straightforward calculation shows that under this circumstances
\bea
\Phi=\Psi=\Omega =0 \; .
\eea
Also, eq. (\ref{eq:realderiv}) implies $ \partial^2_u f{|}_{u_\pm}=0$. One
needs to push the Taylor development around the root
\bea
 u_0 = 1-\xi^2 + \frac{x-y}{2} \; ,
\eea
to third order
\bea
f(u,x,y) = f \Big{|}_{u_0} +  \frac16(u-u_0)^3 [\partial_{u}^3f]\Big{|}_{u_0} + O(u^3)\; ,
\eea
and the Wigner matrix elements has an asymptotic behavior (see \cite{horman}) 
\bea
\int du  \sqrt{K(u,x,y)} e^{J f } \approx e^{J f{|}_{u_0}} 
\Big{\{} \,\mathrm{Ai}(a(x,y)[\imath J]^{\frac23})
[\imath J]^{-\frac{1}{3}}  
 +\,\mathrm{Ai}'(a(x,y)[\imath J]^{\frac23})
[\imath J]^{-\frac23}  \Big{\}} \,,
\label{eq:airy}
\eea
where $a(x,y)$ is some non vanishing smooth real function (determined 
by $K$ and $f$ evaluated at $u_0$, see \cite{horman}), $\mathrm{Ai}$ is
the Airy function of the first kind and $\mathrm{Ai}'$ its derivative. 
At large argument the Airy functions behave like 
\bea
\mathrm{Ai}(\zeta) \approx \frac{e^{-\frac23\zeta^{\frac32}}}{2\sqrt\pi\,\zeta^{\frac14}}
 \approx-\mathrm{Ai}'(\zeta) \;.
\label{airyfun}
\eea
The term $\mathrm{Ai}'$ is therefore subleading and we have
\bea
&& \int du  \sqrt{K(u,x,y)} e^{J f } \approx 
\frac{ e^{\imath J\bigl(\alpha x +\gamma y -\frac23 (a(x,y))^{\frac32}  \bigr) }  }
{ \sqrt{\imath J}(a(x,y))^{1/4}} \; .
\label{asymp}
\eea

\section{Computations for the $3j$ symbol}

In this appendix we detail at length the various computations required for the proof of Theorem \ref{th:3j}.

\subsection{The first derivative}
\label{app:3jderiv}

To compute the derivative $\partial_{\xi^2} \sum_i s_i (\imath  f_i)$, note that 
$ \partial_{(\xi^2)} \Delta_i  = -(2 \xi^2 -1 -x_i y_i)$. The partial derivative of $\imath \phi_i$ is then  
\bea
&&\imath \partial_{(\xi^2)}  \phi_i =  
\frac{2 + 2 \imath \frac{\partial_{(\xi^2)} \Delta_i }{2\sqrt{\Delta_i}}}
{( 2\xi^2 -1- x_i y_i + 2 \imath \sqrt{\Delta_i})} =
\frac{\bigl(2 -  \imath \frac{2 \xi^2 -1 -x_i y_i}{\sqrt{\Delta_i}} \bigr)
\bigl(  2\xi^2 -1- x_i y_i - 2 \imath \sqrt{\Delta_i} \bigr)}
{ (2\xi^2 -1- x_i y_i)^2 + 4 \Delta_i } \crcr
&& = \frac{-\imath }{\sqrt{\Delta_i}}\;,
\eea
while the derivative of $\imath \psi_i$ writes
\bea
\label{eq:ipartpsi}
&& \imath \partial_{(\xi^2)} \psi_i 
= \frac{- x_i  + \imath \frac{\partial_{(\xi^2)} \Delta_i }{2\sqrt{\Delta_i}} }
{\frac{x_i+y_i}{2} - x_i\xi^2 + \imath\sqrt{\Delta_i} } -  \frac{1-2\xi^2}{2\xi^2(1-\xi^2)} \crcr
&&= \imath \frac{
\Big{[} - (2 \xi^2 -1 -x_i y_i) + \imath 2 x_i \sqrt{\Delta_i} \Big{]} 
\Big{[}\frac{x_i+y_i}{2} - x_i\xi^2 - \imath\sqrt{\Delta_i}\Big{]} }
{ 2\sqrt{\Delta_i} \xi^2 (1-\xi^2)(1-x_i^2) } -  \frac{1-2\xi^2}{2\xi^2(1-\xi^2)} \; .
\eea
We first evaluate $2 x_i \Delta_i - \bigl( 2\xi^2-1 -x_iy_i \bigr) \bigl(\frac{x_i+y_i}{2} - x_i\xi^2 \bigr)$ as
\bea 
&&=  2 x_i\Big{[}  -\xi^4 + \xi^2 (1 + x_i y_i) -\frac{(x_i+y_i)^2}{4} \Big{]} 
-\Bigl[ 2 \xi^2 - 1 - x_iy_i \Bigr] \bigl(\frac{x_i+y_i}{2} - x_i\xi^2 \bigr)
\crcr
&&= \xi^2 \Bigl[ x_i( 1+ x_iy_i) -x_i-y_i \Bigr] - \frac{x_i+y_i}{2} 
\Bigl[ x_i (x_i+y_i ) -1- x_iy_i \Bigr]
\crcr
&&=(1-x_i^2) (\frac{x_i+ y_i}{2} - \xi^2y_i) \;,
\eea
hence eq. (\ref{eq:ipartpsi}) writes
\bea 
&&  \imath \frac{
(1-x_i^2) (\frac{x_i+ y_i}{2} - \xi^2y_i) }
{2\sqrt{\Delta_i} \xi^2 (1-\xi^2) (1-x_i^2)  }
+ \imath \frac{ \imath \sqrt{\Delta_i} 
\Big{[} x_i^2 - 2 x_i^2\xi^2 +  2 \xi^2 -1 \Big{]}}
{2\sqrt{\Delta_i} \xi^2 (1-\xi^2) (1-x_i^2)  }
 -  \frac{(1-2\xi^2) }{2 \xi^2(1-\xi^2)}\crcr
&& = \imath \frac{(\frac{x_i+ y_i}{2} - \xi^2y_i)}{2\sqrt{\Delta_i}\xi^2} \; .
\eea
Noting that $\omega (x_i,y_i) = \psi(-y_i,-x_i)$ the derivative of $\imath \omega$ writes 
\bea
\imath \partial_{(\xi^2)} \omega_i = \frac{  
\imath (-\frac{x_i+ y_i}{2} + \xi^2 x_i) }{2\sqrt{\Delta_i}\xi^2(1-\xi^2)} \;.
\eea
The derivative of $\sum_i s_i (\imath f_i)$ is then 
\bea
&& \partial_{ (\xi^2) } \sum_i s_i (\imath f_i) =  \imath \sum_i  s_i J_i 
\Big{(} \frac{-1 }{\sqrt{\Delta_i}} +  x_i
\frac{  (\frac{x_i+ y_i}{2} - \xi^2y_i) }{2\sqrt{\Delta_i}\xi^2(1-\xi^2)}
- y_i \frac{  
(-\frac{x_i+ y_i}{2} + \xi^2 x_i) }{2\sqrt{\Delta_i}\xi^2(1-\xi^2)}
\Big{)} \crcr
&&= \imath \sum_i  s_i J_i 
\Big{(} 
\frac{ -2\xi^2(1-\xi^2) + \frac{(x_i+ y_i)^2}{2} -2\xi^2 x_i y_i }{2\sqrt{\Delta_i}\xi^2(1-\xi^2)}
\Big{)} \crcr
&&= \imath \sum_i  s_i J_i 
\Big{(} 
-2\frac{ \Delta_i }{2\sqrt{\Delta_i}\xi^2(1-\xi^2)}
\Big{)} =
-\frac{\imath}{\xi^2(1-\xi^2)} \sum_i s_iJ_i \sqrt{\Delta_i} \; .
\eea

\subsection{The saddle point equation}
\label{app:3jsaddle}

We will use in the sequel the short hand notation $A^{\vec B}:= \vec A \cdot \vec B$ for 
all vectors $\vec A$ and $\vec B$. Squaring twice the saddle point eq. (\ref{eq:3jsaddleready}) 
we obtain, for all signs $s_i$,
\bea\label{eq:saddleini}
\Big{[} J_3^2 \Delta_3 -J_1^2 \Delta_1 - J_2^2\Delta_2 \Big{]}^2= 4 J_1^2 J_2^2 \Delta_1 \Delta_2 \; .
\eea
We first translate eq. (\ref{eq:saddleini}) in terms of angular momentum vectors
\bea
 J_i^2 \Delta_i = (1-\xi^2) \xi^2 J_i^2 + \xi^2 J_i^{\vec n} J_i^{\vec k} -\frac{1}{4} (J_i^{\vec n+\vec k})^2 \; .
\eea
Then $J_3^2 \Delta_3 -J_1^2 \Delta_1 - J_2^2\Delta_2 $ computes to 
\bea
&& (1-\xi^2) \xi^2 \Bigl[J_3^2 -J_1^2-J_2^2 \Bigr] 
+ \xi^2 \Bigl[ J_3^{\vec n} J_3^{\vec k} - J_1^{\vec n} J_1^{\vec k}-J_2^{\vec n} J_2^{\vec k}\Bigr] \crcr
&& -\frac{1}{4} \Bigl[ (J_3^{\vec n+\vec k})^2 - (J_1^{\vec n+\vec k})^2- (J_2^{\vec n+\vec k})^2 \Bigr] \; ,
\eea
and using $\vec J_3 = -\vec J_1 -\vec J_2$, eq. (\ref{eq:saddleini}) becomes
\bea
&&\Big{\{} 2 (1-\xi^2) \xi^2 \vec J_1 \cdot \vec J_2 + 
\xi^2 \big{(}  J_1^{\vec n} J_2^{\vec k} + J_2^{\vec n} J_1^{\vec k}\big{)}
-\frac{1}{2} J_1^{\vec n +\vec k} J_2^{\vec n +\vec k} \Big{\}}^2 \crcr
&&=  \Big{[} 2 (1-\xi^2) \xi^2 J_1^2 + 2\xi^2 J_1^{\vec n} J_1^{\vec k} -\frac{1}{2} (J_1^{\vec n+\vec k})^2\Big{]} 
\crcr
&&\times \Big{[} 2 (1-\xi^2) \xi^2 J_2^2 + 2\xi^2 J_2^{\vec n} J_2^{\vec k} -\frac{1}{2} (J_2^{\vec n+\vec k})^2\Big{]} \; .
\eea
Collecting all terms on the LHS we get
\bea\label{eq:intermed}
&& 4 (1-\xi^2)^2 \xi^4 \Bigl[ J_1^2J_2^2 - (\vec J_1 \cdot \vec J_2)^2 \Bigr]
\crcr
&&+ 4(1-\xi^2) \xi^4 \Bigl[ J_1^2 J_2^{\vec n} J_2^{\vec k} + J_2^2 J_1^{\vec n} J_1^{\vec k} 
-\vec J_1 \cdot \vec J_2  \big{(}  J_1^{\vec n} J_2^{\vec k} + J_2^{\vec n} J_1^{\vec k}\big{)} \Bigr] \crcr
&&- (1-\xi^2) \xi^2 \Big{[}J_1^2  (J_2^{\vec n+\vec k})^2 + J_2^2  (J_1^{\vec n+\vec k})^2
-2 \vec J_1\cdot \vec J_2 J_1^{\vec n +\vec k} J_2^{\vec n +\vec k} 
\Big{]} \crcr
&&+ \xi^4 \Big{[} 4 J_1^{\vec n} J_1^{\vec k} J_2^{\vec n} J_2^{\vec k} 
-\big{(}J_1^{\vec n} J_2^{\vec k} + J_2^{\vec n} J_1^{\vec k}\big{)}^2  \Big{]} \crcr
&&-\xi^2 \Big{[} J_1^{\vec n} J_1^{\vec k} (J_2^{\vec n+\vec k})^2 + J_2^{\vec n} J_2^{\vec k} (J_1^{\vec n+\vec k})^2
- \big{(}  J_1^{\vec n} J_2^{\vec k} + J_2^{\vec n} J_1^{\vec k}\big{)} J_1^{\vec n +\vec k} J_2^{\vec n +\vec k}
\Big{]} =0 \; .
\eea

Eq. (\ref{eq:intermed}) rewrites
\bea
&& 4 (1-\xi^2)^2 \xi^4 \Bigl[ \vec J_1 \wedge \vec J_2 \Bigr]^2
+ 4(1-\xi^2) \xi^4 \Big{[} \vec n \wedge (\vec J_1 \wedge \vec J_2) \Big{]} 
\cdot \Big{[} \vec k \wedge (\vec J_1 \wedge \vec J_2) \Big{]} \crcr
&&- (1-\xi^2) \xi^2 \Big{[}(\vec n+\vec k) \wedge (\vec J_1 \wedge \vec J_2) \Big{]}^2 
- \xi^4 \Big{[} (\vec n \wedge \vec k) \cdot (\vec J_1 \wedge \vec J_2) \Big{]}^2 \crcr
&&-\xi^2 \Big{[}\vec n \cdot \bigl[ (\vec n +\vec k ) \wedge (\vec J_1 \wedge \vec J_2) \bigr] \Big{]} 
\Big{[} \vec k \cdot \bigl[ (\vec n +\vec k ) \wedge (\vec J_1 \wedge \vec J_2) \bigr] \Big{]} =0.
\eea
Using $\vec S =\vec J_1\wedge \vec J_2$, twice the oriented area of the triangle $\{ \vec J_i \}$, the saddle point
equation becomes
\bea
&&0= 4 (1-\xi^2)^2 \xi^4 S^2 + 4(1-\xi^2) \xi^4 \Big{[} \vec n \wedge \vec S \Big{]} 
\cdot \Big{[} \vec k \wedge \vec S \Big{]} \\
&&- (1-\xi^2) \xi^2 \Big{[}(\vec n+\vec k) \wedge \vec S \Big{]}^2 
- \xi^4 \Big{[} (\vec n \wedge \vec k) \cdot \vec S \Big{]}^2 -\xi^2 \Big{[} \vec S \cdot (\vec n  \wedge \vec k ) \Big{]} 
\Big{[} \vec S \cdot (\vec k  \wedge \vec n )  \Big{]}  \nonumber\; \; ,
\eea
and dividing by $(1-\xi^2)\xi^2$ we obtain
\bea
0 = 4 (1-\xi^2) \xi^2 S^2 + 4 \xi^2 \Big{[} \vec n \wedge \vec S \Big{]} 
\cdot \Big{[} \vec k \wedge \vec S \Big{]} 
+  \Big{[} \vec S \cdot (\vec n \wedge \vec k)  \Big{]}^2 -
\Big{[}(\vec n+\vec k) \wedge \vec S \Big{]}^2  \; ,
\eea
that is
\bea \label{eq:treaba}
&& 0 = 4 \xi^4 S^2 - 4 \xi^2 \Big{[}S^2 + (\vec n \cdot \vec k) S^2 - S^{\vec n} S^{\vec k} \Big{]}
\crcr
&&- S^2 (\vec n \wedge \vec k)^2 + \bigl[ \vec S \wedge (\vec n \wedge \vec k) \bigr]^2
+S^2 (\vec n +\vec k)^2 - (S^{\vec n} +S^{\vec k})^2 \; .
\eea
The last line in eq. (\ref{eq:treaba}) computes
\bea 
&&-S^2 +S^2 (\vec n \cdot \vec k)^2 + (S^{\vec n})^2 +(S^{\vec k})^2 - 2 (\vec n \cdot \vec k) S^{\vec n} S^{\vec k}
\crcr
&&+2S^2 +2S^2 (\vec n \cdot \vec k)- (S^{\vec n})^2 - (S^{\vec k})^2 - 2 S^{\vec n} S^{\vec k} \crcr
&& = [1+ (\vec n \cdot \vec k)]^2S^2 - 2[1+(\vec n \cdot \vec k)] S^{\vec n} S^{\vec k} \; ,
\eea
and eq. (\ref{eq:3jsaddlefinal}) follows.

\subsection{Evaluation of $J_i^2 \Delta_i^{\pm}$}

\label{app:3jJDelta}

Recall that $J_i^2\Delta_i$ is 
\bea \label{eq:deltxi}
 J_i^2 \Delta_i = (1-\xi^2) \xi^2 J_i^2 + \xi^2 J_i^{\vec n} J_i^{\vec k} -\frac{1}{4} (J_i^{\vec n+\vec k})^2 \; .
\eea
Evaluated for $\xi^2_+= \frac{1+(\vec n \cdot \vec k)}{2}$, eq. (\ref{eq:deltxi}) becomes
\bea\label{eq:c1}
 J_i^2 \Delta_i^+ = \frac{1 - (\vec n \cdot \vec k)^2}{4} J_i^2+ \frac{1+(\vec n \cdot \vec k)}{2}
J_i^{\vec n} J_i^{\vec k} -\frac{1}{4} (J_i^{\vec n}+J_i^{\vec k})^2 \; ,
\eea
which further simplifies to
\bea
&& J_i^2 \Delta_i^+ = \frac{1}{4}\Big{\{} (\vec n \wedge \vec k)^2J_i^2 + 2(\vec n \cdot \vec k)J_i^{\vec n} J_i^{\vec k} 
- (J_i^{\vec n})^2-(J_i^{\vec k})^2 \Big{\}} \crcr
&&=\frac{1}{4}\Big{[} (\vec n \wedge \vec k)^2J_i^2 + 
J_i^{\vec n} \bigl[(\vec n\wedge \vec k ) \cdot (\vec k \wedge \vec J_i) \bigr]
 - J_i^{\vec k}  \bigl[(\vec n\wedge \vec k ) \cdot (\vec n \wedge \vec J_i) \bigr]\Big{]} \; ,
\eea
and combining the last two terms this is
\bea
&&\frac{1}{4}\Big{\{} (\vec n \wedge \vec k)^2J_i^2 + 
(\vec n\wedge \vec k ) \cdot \Big{[}\bigl(\vec J_i \wedge(\vec k \wedge \vec n) \bigr) \wedge \vec J_i \Big{]} \Big{\}} \crcr
&&=\frac{1}{4}\Big{\{} (\vec n \wedge \vec k)^2J_i^2 +
(\vec n\wedge \vec k ) \cdot \Big{[} \vec J_i \bigl(\vec J_i \cdot (\vec n \wedge \vec k) \bigr) - (\vec n \wedge \vec k)
 \vec J_i^2 \Big{]} \Big{\}} \; ,
\eea
hence for $\xi^2_+$ we get 
\bea
 J_i \Delta_i^+= \frac{1}{4} \Big{[} \vec J_i \cdot (\vec n \wedge \vec k) \Big{]}^2 \; .
\eea

Evaluated in $\xi^2_-= \frac{1+(\vec n \cdot \vec k)}{2} -\frac{  S^{\vec n}  S^{\vec k}}{S^2}$,
 $J_i^2 \Delta_i$ writes
\bea\label{eq:c2}
 J_i \Delta^-_i&=& \Big{(}\frac{1-(\vec n \cdot \vec k)}{2} + \frac{  S^{\vec n}  S^{\vec k}}{S^2} \Big{)}
\Big{(}\frac{1+(\vec n \cdot \vec k)}{2} -\frac{  S^{\vec n}  S^{\vec k}}{S^2} \Big{)} J_i^2 \crcr
&&+ 
\Big{(}\frac{1+(\vec n \cdot \vec k)}{2} -\frac{  S^{\vec n}  S^{\vec k}}{S^2} \Big{)}
J_i^{\vec n} J_i^{\vec k} -\frac{1}{4} (J_i^{\vec n+\vec k})^2 \; .
\eea
Combining all the terms common to the RHS in eq. (\ref{eq:c1}) and eq. (\ref{eq:c2}), we get 
\bea
 J_i \Delta^-_i&=&\frac{1}{4} \Big{[} \vec J_i \cdot (\vec n \wedge \vec k) \Big{]}^2 + 
\frac{S^{\vec n}S^{\vec k}}{S^2} \Big{[} (\vec n \cdot \vec k)J_i^2 - J_i^{\vec n} J_i^{\vec k} \Big{]}
- J_i^2 \frac{  ( S^{\vec n}  S^{\vec k} )^2}{S^4} \\
&=&\frac{1}{4S^2} \Big{\{} \Bigl[ \vec S \bigl( \vec J_i \cdot (\vec n \wedge \vec k)  \bigr)  \Bigr]^2 
+ 4 S^{\vec n}S^{\vec k} \Big{[} (\vec n \wedge \vec J_i ) \cdot (\vec k \wedge \vec  J_i)  \Big{]}
- 4 J_i^2 \frac{( S^{\vec n}  S^{\vec k} )^2}{S^2}  \Big{\}} \nonumber \; .
\eea
But note that $\vec S \cdot \vec J_i =0$, hence the first term on the RHS above can be written as a double vector product,
that is
\bea
 J_i \Delta^-_i&=&\frac{1}{4S^2} \Big{\{} \Bigl[ \vec J_i \wedge  \bigl( \vec S \wedge (\vec n \wedge \vec k)  \bigr)  \Bigr]^2 
+ 4 S^{\vec n}S^{\vec k} \Big{[}   (\vec n \wedge \vec J_i ) \cdot (\vec k \wedge \vec  J_i)      \Big{]}
- 4 J_i^2 \frac{( S^{\vec n}  S^{\vec k} )^2}{S^2}  \Big{\}} \crcr
&=& \frac{1}{4S^2} \Big{\{} \bigl[ \vec J_i \wedge \bigl( \vec n  S^{\vec k} + \vec k S^{\vec n}  \bigr) \bigr]^2
- 4 J_i^2 \frac{( S^{\vec n}  S^{\vec k} )^2}{S^2} 
 \Big{\}} \crcr
&&=  \frac{1}{4S^4} \Big{\{}  S^2 \bigl[ \vec J_i \wedge \bigl( \vec n  S^{\vec k} + \vec k S^{\vec n}  \bigr) \bigr]^2
- 4 J_i^2 ( S^{\vec n}  S^{\vec k} )^2 \Big{\}} \; .
\eea
And, as $A^2B^2 = (\vec A \cdot \vec B)^2 + (\vec A \wedge \vec B)^2$, we have
\bea
&& J_i \Delta^-_i=\frac{1}{4S^4} \Big{\{} 
\Big{[} \vec S \cdot \bigl[ \vec J_i \wedge \bigl( \vec n  S^{\vec k} + \vec k S^{\vec n}  \bigr) \bigr] \Big{]}^2 
+ \bigl[ \vec J_i (S^{\vec n }S^{\vec k} + S^{\vec k} S^{\vec n})\bigr]^2 -   4 J_i^2 ( S^{\vec n}  S^{\vec k} )^2 
\Big{\}} \crcr
&&= \frac{ \Big{\{} \vec J_i \cdot \bigr[ (\vec n \wedge \vec S) S^{\vec k}+
(\vec k \wedge \vec S) S^{\vec n} \bigl]
 \Big{\}}^2}{4 S^4}
=\frac{ \Big{\{} \vec J_i \cdot \bigr[ (\vec S \wedge \vec n) S^{\vec k}+
(\vec S \wedge \vec k) S^{\vec n} \bigl]
 \Big{\}}^2}{4 S^4}
 \; .
\eea

\subsection{Second derivative}
\label{app:3jsecondderiv}

Using $J_i \sqrt{\Delta_i^+}$ from eq. (\ref{eq:signs}) and $\xi^2_+$, we have
\bea
 \epsilon^{+}_2 \epsilon^{+}_3 
J_2\sqrt{\Delta_2^{+}} J_3\sqrt{\Delta_3^{+}} 
\Bigl[ (2 \xi_{+}^2 -1 )J_1^2 -  J_1^{\vec n}  J_1^{\vec k} \Bigr] 
+ \circlearrowleft_{123} &=&\crcr
= 
 \frac{1}{4} \Big{\{} 
J_2^{\vec n \wedge \vec k} J_3^{\vec n \wedge \vec k} [(\vec n \wedge \vec J_1) \cdot(\vec k \wedge \vec J_1)] 
 &+&J_3^{\vec n \wedge \vec k} J_1^{\vec n \wedge \vec k} [(\vec n \wedge \vec J_2) \cdot(\vec k \wedge \vec J_2)] \crcr
 +J_1^{\vec n \wedge \vec k} J_2^{\vec n \wedge \vec k} [(\vec n \wedge \vec J_3) \cdot(\vec k \wedge \vec J_3)]
\Big{\}} \; . &&
\eea
Substituting in the equation above $\vec J_3 = -\vec J_1 - \vec J_2$, the RHS writes
\bea
&&\frac{1}{4} \Big{\{} 
-J_2^{\vec n \wedge \vec k} J_2^{\vec n \wedge \vec k} [(\vec n \wedge \vec J_1) \cdot(\vec k \wedge \vec J_1)]
-J_2^{\vec n \wedge \vec k} J_1^{\vec n \wedge \vec k} [(\vec n \wedge \vec J_1) \cdot(\vec k \wedge \vec J_1)] \crcr
&& -J_1^{\vec n \wedge \vec k} J_1^{\vec n \wedge \vec k} [(\vec n \wedge \vec J_2) \cdot(\vec k \wedge \vec J_2)]
-J_2^{\vec n \wedge \vec k} J_1^{\vec n \wedge \vec k} [(\vec n \wedge \vec J_2) \cdot(\vec k \wedge \vec J_2)]\crcr
&& +J_1^{\vec n \wedge \vec k} J_2^{\vec n \wedge \vec k} \Big{[}(\vec n \wedge \vec J_1) \cdot(\vec k \wedge \vec J_1)+
(\vec n \wedge \vec J_1) \cdot(\vec k \wedge \vec J_2) \crcr
&&+(\vec n \wedge \vec J_2) \cdot(\vec k \wedge \vec J_1) +(\vec n \wedge \vec J_2) \cdot(\vec k \wedge \vec J_2) \Big{]}
\Big{\}} \; ,
\eea
canceling the appropriate cross terms, the remaining expression factors as 
\bea
&&-\frac{1}{4} \Big{\{} \Big{[} J_2^{\vec n \wedge \vec k} (\vec n \wedge \vec J_1) 
-  J_1^{\vec n \wedge \vec k} (\vec n \wedge \vec J_2) \Big{]}
\cdot \Big{[}J_2^{\vec n \wedge \vec k} (\vec k \wedge \vec J_1)
 -J_1^{\vec n \wedge \vec k} (\vec k \wedge \vec J_2)  \Big{]} \Big{\}} \crcr
&&=-\frac{1}{4} \Big{\{} \vec n \wedge \Big{(}(\vec n \wedge \vec k) \wedge (\vec J_1 \wedge \vec J_2) \Big{)} \Big{\} }
\cdot 
\Big{\{} \vec k \wedge \Big{(}(\vec n \wedge \vec k) \wedge (\vec J_1 \wedge \vec J_2) \Big{)} \Big{\} } \; ,
\eea
developing the double vector products and taking into account that 
$ \vec n \cdot (\vec n \wedge \vec k)=\vec k \cdot (\vec n \wedge \vec k) = 0 $, we conclude 
\bea
 \epsilon^{+}_2 \epsilon^{+}_3 
J_2\sqrt{\Delta_2^{+}} J_3\sqrt{\Delta_3^{+}} 
\Bigl[ (2 \xi_{+}^2 -1 )J_1^2 -  J_1^{\vec n}  J_1^{\vec k} \Bigr] 
+ \circlearrowleft_{123} =  -\frac{1}{4}  S^{\vec n} S^{\vec k}  (\vec n\wedge \vec k)^2 \; .
\eea

For the $\xi^2_-$ root we have
\bea\label{eq:3jsecondxi-}
&&  \epsilon^{-}_2 \epsilon^{-}_3
J_2\sqrt{\Delta_2} J_3\sqrt{\Delta_3} \Bigl[ (2 \xi_{-}^2 -1 )J_1^2 - (\vec n \cdot \vec J_1) (\vec k \cdot \vec J_1)\Bigr] 
+ \circlearrowleft_{123}=\\
&& = \frac{1}{4} \frac{\Big{[}  J_2^{\vec S \wedge \vec n}  S^{\vec k} +  
 J_2^{\vec S \wedge \vec k} S^{\vec n}
\Big{]}}{S^2}
\frac{\Big{[}  J_3^{\vec S \wedge \vec n} S^{\vec k} +  
 J_3^{\vec S \wedge \vec k} S^{\vec n}
\Big{]}}{S^2} \Big{[}(\vec n \wedge \vec J_1) \cdot(\vec k \wedge \vec J_1) 
- 2 J_1^2\frac{ S^{\vec n} S^{\vec k }}{S^2} \Big{]}
\crcr
&&+\frac{1}{4} \frac{\Big{[}  J_1^{\vec S \wedge \vec n} S^{\vec k} +  
 J_1^{\vec S \wedge \vec k} S^{\vec n}
\Big{]}}{S^2}
\frac{\Big{[}  J_3^{\vec S \wedge \vec n} S^{\vec k} +  
 J_3^{\vec S \wedge \vec k} S^{\vec n}
\Big{]}}{S^2} \Big{[}(\vec n \wedge \vec J_2) \cdot(\vec k \wedge \vec J_2) 
- 2J_2^2\frac{ S^{\vec n} S^{\vec k}}{S^2} \Big{]}
\crcr
&&+\frac{1}{4} \frac{\Big{[}  J_1^{\vec S \wedge \vec n} S^{\vec k}+  
 J_1^{\vec S \wedge \vec k} S^{\vec n}
\Big{]}}{S^2}
\frac{\Big{[}  J_2^{\vec S \wedge \vec n} S^{\vec k} +  
 J_2^{\vec S \wedge \vec k} S^{\vec n}
\Big{]}}{S^2} \Big{[}(\vec n \wedge \vec J_3) \cdot(\vec k \wedge \vec J_3) 
- 2J_3^2\frac{ S^{\vec n} S^{\vec k} }{S^2} \Big{]} \; .
\nonumber
\eea
We substitute again in the equation above $\vec J_3 = -\vec J_1-\vec J_2$. The coefficient of 
$\frac{1}{4S^2}$ computes, canceling the appropriate cross terms,
\bea 
&&-\Big{[}  J_2^{\vec S \wedge \vec n} S^{\vec k} + J_2^{\vec S \wedge \vec k} S^{\vec n}  \Big{]}^2 
(\vec n \wedge \vec J_1) \cdot (\vec k \wedge \vec J_1)  
-\Big{[}  J_1^{\vec S \wedge \vec n} S^{\vec k} + J_1^{\vec S \wedge \vec k} S^{\vec n}  \Big{]}^2 
(\vec n \wedge \vec J_2) \cdot (\vec k \wedge \vec J_2) \nonumber\\
&&+ \Big{[}  J_1^{\vec S \wedge \vec n} S^{\vec k} + J_1^{\vec S \wedge \vec k} S^{\vec n}  \Big{]}
\Big{[}J_2^{\vec S \wedge \vec n} S^{\vec k} + J_2^{\vec S \wedge \vec k} S^{\vec n}  \Big{]} 
\nonumber\\
&&\times \Big{[} (\vec n \wedge \vec J_1) \cdot (\vec k \wedge \vec J_2)
+(\vec n \wedge \vec J_2) \cdot (\vec k \wedge \vec J_1)     \Big{]}
 \; , 
\eea 
while the coefficient of $-\frac{S^{\vec k} S^{\vec n}}{2S^6}$ is
\bea
&& -  J_1^2 \Big{[}  J_2^{\vec S \wedge \vec n} S^{\vec k} + J_2^{\vec S \wedge \vec k} S^{\vec n}  \Big{]}^2  
 -  J_2^2 \Big{[}  J_1^{\vec S \wedge \vec n} S^{\vec k} + J_1^{\vec S \wedge \vec k} S^{\vec n}  \Big{]}^2 \nonumber\\
&& +2 \vec J_1\cdot \vec J_2 \Big{[}  J_1^{\vec S \wedge \vec n} S^{\vec k} + J_1^{\vec S \wedge \vec k} S^{\vec n}  \Big{]} 
 \Big{[}J_2^{\vec S \wedge \vec n} S^{\vec k} + J_2^{\vec S \wedge \vec k} S^{\vec n}  \Big{]} \; .
\eea
The RHS of eq. (\ref{eq:3jsecondxi-}) becomes 
\bea
 \frac{-1}{4S^4} 
&&\Big{[}  \Big{(} J_2^{\vec S \wedge \vec n} S^{\vec k} + J_2^{\vec S \wedge \vec k} S^{\vec n}  \Big{)}
(\vec n \wedge \vec J_1) 
-  \Big{(}  J_1^{\vec S \wedge \vec n} S^{\vec k} + J_1^{\vec S \wedge \vec k} S^{\vec n}  \Big{)}
(\vec n \wedge \vec J_2) 
\Big{]}     \crcr
&& \cdot 
\Big{[} \Big{(} J_2^{\vec S \wedge \vec n} S^{\vec k} + J_2^{\vec S \wedge \vec k} S^{\vec n}   \Big{)}
(\vec k \wedge \vec J_1) 
- \Big{(}  J_1^{\vec S \wedge \vec n} S^{\vec k} + J_1^{\vec S \wedge \vec k} S^{\vec n}  \Big{)}
(\vec k \wedge \vec J_2) 
\Big{]} 
\crcr 
+\frac{S^{\vec k} S^{\vec n}}{2S^6} && \Big{[} \vec J_1  
\Big{(}  J_2^{\vec S \wedge \vec n} S^{\vec k} + J_2^{\vec S \wedge \vec k} S^{\vec n}  \Big{)} 
- \vec J_2 \Big{(}  J_1^{\vec S \wedge \vec n} S^{\vec k} + J_1^{\vec S \wedge \vec k} S^{\vec n}  \Big{)} 
\Big{]}^2 \; ,
\eea
which rewrites, combining the appropriate terms into double vector products as
\bea
 \frac{-1}{4S^4} 
&&\Big{\{} \vec n \wedge \Big{[}   (\vec S \wedge \vec n) \wedge ( \vec J_1\wedge \vec J_2 )  S^{\vec k}
+  (\vec S \wedge \vec k) \wedge ( \vec J_1\wedge \vec J_2 ) S^{\vec n}
  \Big{]}
\Big{\} }     \crcr
&& \cdot 
\Big{\{} \vec k \wedge \Big{[}   (\vec S \wedge \vec n) \wedge ( \vec J_1\wedge \vec J_2 )  S^{\vec k}
+  (\vec S \wedge \vec k) \wedge ( \vec J_1\wedge \vec J_2 ) S^{\vec n}
  \Big{]}
\Big{\} } 
\crcr 
+\frac{S^{\vec k} S^{\vec n}}{2S^6} && \Big{[} (\vec S \wedge \vec n) \wedge ( \vec J_1\wedge \vec J_2 )  S^{\vec k}
+  (\vec S \wedge \vec k) \wedge ( \vec J_1\wedge \vec J_2 ) S^{\vec n}
\Big{]}^2 \; ,
\eea
and recalling that $\vec J_1 \wedge \vec J_2 = \vec S$ the above equation becomes 
\bea
 \frac{-1}{4S^4} 
&&\Big{\{} \vec n \wedge \Big{[}   (\vec S \wedge \vec n) \wedge \vec S S^{\vec k}
+  (\vec S \wedge \vec k) \wedge \vec S S^{\vec n}
  \Big{]}
\Big{\} }     \crcr
&& \cdot 
\Big{\{} \vec k \wedge \Big{[}   (\vec S \wedge \vec n) \wedge \vec S S^{\vec k}
+  (\vec S \wedge \vec k) \wedge \vec S) S^{\vec n}
  \Big{]}
\Big{\} } 
\crcr 
+\frac{S^{\vec k} S^{\vec n}}{2S^6} && \Big{[}  (\vec S \wedge \vec n) \wedge \vec S S^{\vec k}
+  (\vec S \wedge \vec k) \wedge \vec S S^{\vec n}
\Big{]}^2 \; .
 \eea
We develop the double vector products in the first line and take into account $\vec n \cdot (\vec S \wedge \vec n)= 
 \vec k \cdot (\vec S \wedge \vec k)= 0 $. For the second line we use 
$(\vec S\wedge \vec A)^2= S^2A^2 - (\vec S \cdot \vec A)^2$ and
 $\vec S \cdot (\vec S\wedge \vec n)= \vec S \cdot (\vec S \wedge \vec k) = 0$ to rewrite the equation as
\bea
 \frac{-S^{\vec n} S^{\vec k} }{4S^4} 
&&\Big{\{} - \Big{[}  \vec n \cdot (\vec S \wedge \vec k) \Big{]} \vec S+ (\vec S \wedge \vec n) S^{\vec k} 
+  (\vec S \wedge \vec k) S^{\vec n}
  \Big{]}
\Big{\} }     \crcr
&& \cdot 
\Big{\{} - \Big{[} \vec k \cdot (\vec S \wedge \vec n) \Big{]} \vec S + (\vec S \wedge \vec n) S^{\vec k}
 + (\vec S \wedge \vec k) S^{\vec n} 
  \Big{]}
\Big{\} } 
\crcr 
+\frac{S^{\vec k} S^{\vec n}}{2S^4} && \Big{[} (\vec S \wedge \vec n) S^{\vec k}
+  (\vec S \wedge \vec k) S^{\vec n}
\Big{]}^2 \; ,
\eea
and noting that the cross term in the first scalar product cancel 
(again as $\vec S \cdot (\vec S\wedge \vec n)= \vec S \cdot (\vec S \wedge \vec k) = 0$), and combining the remaining
three terms we get 
\bea
&& \frac{S^{\vec n} S^{\vec k} }{4S^4} S^2 \Big{[} \vec S \cdot (\vec n\wedge \vec k) \Big{]}^2
+\frac{S^{\vec k} S^{\vec n}}{4S^4} \Big{[} (\vec S \wedge \vec n) S^{\vec k}
+  (\vec S \wedge \vec k) S^{\vec n}
\Big{]}^2  \; .
\eea
Factoring $\vec S$ in the second term and using $A^2 B^2 =(\vec A \cdot \vec B)^2+ (\vec A \wedge \vec B)^2$ this 
rewrites as
\bea
&& \frac{S^{\vec n} S^{\vec k}}{4} \Big{[}  (\vec n\wedge \vec k)^2  -
\frac{[\vec n \vec S^{\vec k} - \vec k  \vec S^{\vec n} ]^2 }{S^2}
+\frac{[\vec n S^{\vec k} + \vec k S^{\vec n}]^2 }{S^2} - \frac{4 (S^{\vec n} S^{\vec k})^2}{S^4}\Big{]} \; ,
 \eea
thus we conclude
\bea
&&\epsilon^{-}_2 \epsilon^{-}_3
J_2\sqrt{\Delta_2} J_3\sqrt{\Delta_3} \Bigl[ (2 \xi_{-}^2 -1 )J_1^2 - (\vec n \cdot \vec J_1) (\vec k \cdot \vec J_1)\Bigr] 
+ \circlearrowleft_{123}=\crcr
&&= \frac{S^{\vec n} S^{\vec k}}{4} \Big{[}  (\vec n\wedge \vec k)^2  
+4 (\vec n\cdot \vec k) \frac{S^{\vec n} S^{\vec k} }{S^2} - \frac{4 (S^{\vec n} S^{\vec k})^2}{S^4}\Big{]} \; .
\eea

\subsection{Function at the saddle points}
\label{app:3jangles}

We evaluate the relevant angles at the points $\xi^2_{\pm}$ by substituting 
 eq. (\ref{eq:rootsxi}) and eq. (\ref{eq:signs}) into eq. (\ref{eq:phichiom}).

\subsubsection{The angles $\phi_i^{\pm}$} 

For the angles $\phi_i^{\pm}$ the direct substitution yields
\bea
&&  \imath \epsilon_i^+\phi_i^+=  \ln \frac{ (\vec n\wedge \vec J_i)\cdot(\vec k\wedge \vec J_i) 
 + \imath J_i  \Bigl[\vec J_i \cdot (\vec n\wedge \vec k) \Bigl]
 }
{ \sqrt{(\vec n \wedge \vec J_i)^2} \sqrt{(\vec k \wedge \vec J_i)^2}}
\\
&& \imath \epsilon_i^-\phi_i^-= \ln \frac{ (\vec n\wedge \vec J_i)\cdot(\vec k\wedge \vec J_i) 
 -2 J_i^2\frac{ S^{\vec n} \vec S^{\vec k} }{S^2} 
 + \imath J_i   
\frac{\vec J_i \cdot \Big{[} (\vec S \wedge \vec n)  S^{\vec k}
+  (\vec S \wedge \vec k)   S^{\vec n} \Big{]}}{S^2} 
 }
{ \sqrt{(\vec n \wedge \vec J_i)^2 }  \sqrt{(\vec k \wedge \vec J_i)^2}} \; .\nonumber
\eea

Consider first the denominator of $\imath \phi_i^{-}$ multiplied by $S^2$, namely
\bea
&&S^2 (\vec n\wedge \vec J_i)\cdot(\vec k\wedge \vec J_i) 
 -2 J_i^2 S^{\vec n} S^{\vec k} 
 + \imath J_i \vec J_i \cdot \Big{[} (\vec S \wedge \vec n) S^{\vec k}
+  (\vec S \wedge \vec k) S^{\vec n} \Big{]} \crcr
&&= \bigl[ \vec S \wedge ( \vec n\wedge \vec J_i )\bigr] \cdot \bigl[ \vec S \wedge ( \vec k\wedge \vec J_i )\bigr]
+ [\vec S \cdot (\vec n \wedge \vec J_i) ][ \vec S \cdot (\vec k \wedge \vec J_i) ] \crcr
&&-2 J_i^2 S^{\vec n} S^{\vec k} 
 + \imath J_i \vec J_i \cdot \Big{[} (\vec S \wedge \vec n) S^{\vec k}
+  (\vec S \wedge \vec k) S^{\vec n} \Big{]} \crcr
&&=\Bigl[ \vec n \cdot (\vec J_i \wedge \vec S) + \imath J_i S^{\vec n} \Bigr] 
\Bigl[ \vec k \cdot (\vec J_i \wedge \vec S) + \imath J_i S^{\vec k} \Bigr] \; ,
\eea
hence 
\bea
 \imath \epsilon_i^-\phi_i^-= \imath \Phi^i_{\vec n} +\imath \Phi^i_{\vec k} \qquad 
\imath\Phi^i_{\vec n} = \ln \frac{ \vec n \cdot (\vec J_i \wedge \vec S) + \imath J_i S^{\vec n} }
{S \sqrt{(\vec n \wedge \vec J_i)^2 } } \; .
\eea
Note that 
\bea
&& \Big{[} \vec n \cdot (\vec J_i \wedge \vec S) + \imath J_i S^{\vec n}  \Big{]} 
\Big{[} \vec k \cdot (\vec J_i \wedge \vec S) - \imath J_i S^{\vec k} \Big{]} 
 = [\vec S \cdot (\vec n \wedge \vec J_i) ][ \vec S \cdot (\vec k \wedge \vec J_i) ] 
+ J_i^2 S^{\vec n} S^{\vec k} \crcr
&&+\imath J_i \vec J_i \cdot \Big{[} \vec S \wedge \Big{(} \vec k S^{\vec n} - \vec n 
S^{\vec k } \Big{)} \Big{]} = 
S^2 (\vec n\wedge \vec J_i)\cdot(\vec k\wedge \vec J_i)
-\bigl[ \vec S \wedge ( \vec n\wedge \vec J_i )\bigr] \cdot \bigl[ \vec S \wedge ( \vec k\wedge \vec J_i )\bigr] \crcr
&& + J_i^2 S^{\vec n} S^{\vec k} +\imath J_i \vec J_i \cdot \Big{[} \vec S \wedge \Big{(} \vec S \wedge(\vec k \wedge \vec n )
\Big{)} \Big{]} \; ,
\eea
and developing the double vector products, taking into account $\vec S \cdot \vec J_i = 0$, we conclude 
\bea
 \imath \epsilon_i^+\phi_i^+ = \imath \Phi^i_{\vec n} - \imath \Phi^i_{\vec k} \; .
\eea

\subsubsection{The angles $\epsilon_1^{\pm}\psi_1^{\pm}- \epsilon_3^{\pm}\psi_3^{\pm}$}

We will denote in this section $\vec A \wedge \vec B = \vec A^{\wedge \vec B}$
Direct substitution of $\xi^2_+$ and $\xi^2_-$ yields 
\bea
&& \imath \epsilon_i^+ \psi_i^+=  \ln \frac{ J_i^{\vec k} - 
J_i^{\vec n} (\vec n\cdot \vec k) + \imath \vec J_i\cdot  (\vec n\wedge \vec k)}
{\sqrt{ \bigl[ 1-(\vec n\cdot \vec k)^2 \bigr] (\vec n\wedge \vec J_i)^2 }} 
\\
&& \imath \epsilon_i^- \psi_i^- =  \ln 
\frac{ J_i^{\vec k} - J_i^{\vec n}  (\vec n \cdot \vec k) + 2  J_i^{\vec n} 
  \frac{ S^{\vec n} S^{\vec k}}{S^2} 
+ \imath \frac{\vec J_i \cdot \Big{[} (\vec S \wedge \vec n) S^{\vec k}
+  (\vec S \wedge \vec k)   S^{\vec n} \Big{]}}{S^2}   }
{ \sqrt{ \Big{[} 1- (\vec n\cdot \vec k)^2 +4 (\vec n \cdot \vec k)\frac{S^{\vec n} S^{\vec k}} {S^2} - 
4\frac{ (S^{\vec n} S^{\vec k})^2}{S^4}\Big{]} (\vec n\wedge \vec J_i)^2 } } \; .\nonumber
\eea
To evaluate $\imath \epsilon_1^+ \psi_1^+ - \imath \epsilon_3^+ \psi_3^+ $ we take apart the numerator
\bea
&&\Big{[} J_1^{\vec n\wedge (\vec k\wedge \vec n)} + \imath J_1^{\vec n\wedge \vec k}  \Big{]}
\Big{[} J_3^{\vec n\wedge (\vec k\wedge \vec n)} - \imath J_3^{\vec n\wedge \vec k}  \Big{]}  \crcr
&&=-\vec J_1^{\;\wedge \bigl[\vec n\wedge (\vec k\wedge \vec n)\bigr]} \cdot 
\vec J_3^{\;\wedge \bigl[ \vec n\wedge (\vec k\wedge \vec n)\bigr] } 
+\vec J_1 \cdot \vec J_3 \Big{(} \vec n\wedge (\vec k\wedge \vec n)\Big{)}^2 + 
 J_1^{\vec n\wedge \vec k} J_3^{\vec n\wedge \vec k} 
\crcr
&&+ \imath \Big{(}J_1^{\vec n\wedge \vec k} J_3^{\vec n\wedge (\vec k\wedge \vec n)} -
J_1^{\vec n\wedge (\vec k\wedge \vec n)}J_3^{\vec n\wedge \vec k}
  \Big{)} \; .
\eea
Taking into account $\vec n \cdot (\vec k\wedge \vec n)=0$ this writes
\bea
&& - (\vec k \wedge \vec n)^2 J_1^{\vec n} J_3^{\vec n} + \vec J_1 \cdot \vec J_3 (\vec k\wedge \vec n)^2
+\imath \vec J_1 \cdot \Big{[} \vec J_3 \wedge \Big{[} (\vec n\wedge \vec k) \wedge \bigr( n\wedge (\vec k\wedge \vec n) \bigl) \Big{]}   \Big{]} \crcr
&&= (\vec k \wedge \vec n)^2  (\vec n \wedge \vec J_1)\cdot (\vec n \wedge \vec J_3)  
-\imath \vec J_1 \cdot 
\Big{[} \vec J_3 \wedge \Big{[} \vec n (\vec n\wedge \vec k)^2   \Big{]} \; ,
\eea
hence
\bea
 \imath  \epsilon_1^{+} \psi_1^{+}- \imath  \epsilon_3^{+} \psi_3^{+}  = 
\ln \frac{ (\vec n \wedge \vec J_1)\cdot (\vec n \wedge \vec J_3)-  
\imath \vec n \cdot (\vec J_1 \wedge \vec J_3)}
{\sqrt{ (\vec n\wedge \vec J_1)^2(\vec n\wedge \vec J_3)^2 }  } = \imath \Psi^{13}_{\vec n} \; .
\eea

To evaluate $\imath \epsilon_1^- \psi_1^- - \imath \epsilon_3^- \psi_3^- $ we again take apart the numerator
\bea
&& \Big{(} J_1^{\vec n \wedge (\vec k \wedge \vec n) + 2 \vec n  \frac{  S^{\vec n} S^{\vec k} }{S^2}  }
+\imath J_1^{\vec S \wedge \vec n \frac{S^{\vec k}}{S^2} + \vec S \wedge \vec k \frac{S^{\vec n}}{S^2} }\Big{)}
\Big{(} J_3^{\vec n \wedge (\vec k \wedge \vec n) + 2 \vec n  \frac{  S^{\vec n} S^{\vec k} }{S^2}  }
- \imath J_3^{\vec S \wedge \vec n \frac{S^{\vec k}}{S^2} + \vec S \wedge \vec k \frac{S^{\vec n}}{S^2} }\Big{)} 
\; .
\label{eq:numera}
\eea

The real part is
\bea
&& J_1^{\vec n \wedge (\vec k \wedge \vec n) + 2 \vec n  \frac{  S^{\vec n} S^{\vec k} }{S^2}  }
J_3^{\vec n \wedge (\vec k \wedge \vec n) + 2 \vec n  \frac{  S^{\vec n} S^{\vec k} }{S^2}  }
+J_1^{\vec S \wedge \vec n \frac{S^{\vec k}}{S^2} + \vec S \wedge \vec k \frac{S^{\vec n}}{S^2} }
J_3^{\vec S \wedge \vec n \frac{S^{\vec k}}{S^2} + \vec S \wedge \vec k \frac{S^{\vec n}}{S^2} } \crcr
&&= -\vec J_1^{\;\wedge \bigl[ \vec n \wedge (\vec k \wedge \vec n) + 2 \vec n  \frac{  S^{\vec n} S^{\vec k} }{S^2} 
\bigr] }
\cdot
\vec J_3^{\;\wedge \bigl[ \vec n \wedge (\vec k \wedge \vec n) + 2 \vec n  \frac{  S^{\vec n} S^{\vec k} }{S^2} 
\bigr] }
+\vec J_1 \cdot \vec J_3 \Big{(} \vec n \wedge (\vec k \wedge \vec n) + 2 \vec n  \frac{  S^{\vec n} S^{\vec k} }{S^2} \Big{)}^2 \crcr
&& - \vec J_1^{\;\wedge \bigl[ \vec S \wedge \vec n \frac{S^{\vec k}}{S^2} + \vec S \wedge \vec k \frac{S^{\vec n}}{S^2} \bigr] }
\cdot
\vec J_3^{\;\wedge \bigl[ \vec S \wedge \vec n \frac{S^{\vec k}}{S^2} + \vec S \wedge \vec k \frac{S^{\vec n}}{S^2}
\bigr] } 
+ \vec J_1 \cdot \vec J_3 \Big{(}  \vec S \wedge \vec n \frac{S^{\vec k}}{S^2} + \vec S \wedge \vec k \frac{S^{\vec n}}{S^2}  \Big{)}^2 \; ,
\eea
which rewrites, taking into account $\vec S \cdot \vec J_i =0 $,
\bea
&&-\Big{(} \vec n J_1^{\vec k\wedge \vec n} - (\vec k \wedge \vec n)  J_1^{\vec n} +
 2 \vec J_1\wedge \vec n \frac{  S^{\vec n} S^{\vec k} }{S^2}  \Big{)} \cdot
\Big{(} \vec n J_3^{\vec k\wedge \vec n} - (\vec k \wedge \vec n) J_3^{\vec n} +
 2 \vec J_3\wedge \vec n \frac{  S^{\vec n} S^{\vec k} }{S^2}  \Big{)} \crcr
&& - \vec S \Big{(} J_1^{\vec n} \frac{S^{\vec k}}{S^2} +J_1^{\vec k} \frac{S^{\vec n}}{S^2}  \Big{)} 
\cdot \vec S \Big{(} J_3^{\vec n} \frac{S^{\vec k}}{S^2} +J_3^{\vec k} \frac{S^{\vec n}}{S^2} \Big{)} \crcr
&& +\vec J_1 \cdot \vec J_3  \Big{[} [\vec n \wedge (\vec k\wedge \vec n)]^2 + 
4\frac{(S^{\vec n} S^{\vec k})^2}{S^4} + S^2 \frac{(\vec nS^{\vec k}+\vec k S^{\vec n})^2}{S^4}
-4\frac{(S^{\vec n} S^{\vec k})^2}{S^4}
\Big{]} \; .
\eea
Developing the products in the first line we get
\bea
&& -J_1^{\vec k\wedge \vec n}J_3^{\vec k\wedge \vec n} - (\vec n\wedge\vec k)^2J_1^{\vec n}J_3^{\vec n}
-4 \frac{ (S^{\vec n} S^{\vec k})^2}{S^2} (\vec J_1\wedge \vec n)\cdot(\vec J_3\wedge \vec n) \\
&&+2 \frac{S^{\vec n} S^{\vec k}}{S^2} \Big{[}(\vec J_1\wedge \vec n)\cdot (\vec k\wedge \vec n) J_3^{\vec n}+
(\vec J_3\wedge \vec n)\cdot (\vec k\wedge \vec n) J_1^{\vec n}
\Big{]} 
\crcr
&&- \frac{1}{S^2} \Big{(}J_1^{\vec n}S^{\vec k} + J_1^{\vec k}S^{\vec n} \Big{)} 
\Big{(} J_3^{\vec n}S^{\vec k} + J_3^{\vec k}S^{\vec n} \Big{)} + 
\vec J_1 \cdot\vec  J_3 \Big{(} (\vec n \wedge \vec k)^2  + 
\frac{  (\vec nS^{\vec k}+\vec k S^{\vec n})^2  }{S^2}\Big{)} \; ,
\nonumber
\eea
which is, expanding the second line,
\bea
&& (\vec J_1\wedge \vec n)\cdot(\vec J_3\wedge \vec n) \Big{(} (\vec n \wedge \vec k)^2 -4 \frac{ (S^{\vec n} S^{\vec k})^2}{S^2}  \Big{)} 
- J_1^{\vec k\wedge \vec n}J_3^{\vec k\wedge \vec n} \crcr
&&+ 2 \frac{S^{\vec n} S^{\vec k}}{S^2} \Big{[}J_1^{\vec k} J_3^{\vec n} + J_1^{\vec n} J_3^{\vec k}
- 2 (\vec n\cdot \vec k) J_1^{\vec n} J_3^{\vec n}\Big{]} - 
\frac{1}{S^2} \Big{(}J_1^{\vec n}S^{\vec k} + J_1^{\vec k}S^{\vec n} \Big{)} 
\Big{(} J_3^{\vec n}S^{\vec k} + J_3^{\vec k}S^{\vec n} \Big{)} 
\crcr
&&+\vec J_1 \cdot \vec J_3 \frac{(\vec nS^{\vec k}+\vec k S^{\vec n})^2}{S^2} \; .
\eea
Combining the cross terms in the second line and using $J_1^{\vec k\wedge \vec n}J_3^{\vec k\wedge \vec n} =
J_1^{\vec n\wedge \vec k}J_3^{\vec n\wedge \vec k} $ we obtain 
\bea
&&  (\vec J_1\wedge \vec n)\cdot(\vec J_3\wedge \vec n) \Big{(} (\vec n \wedge \vec k)^2 -4 \frac{ (S^{\vec n} S^{\vec k})^2}{S^2}  \Big{)} 
-4(\vec n\cdot \vec k) J_1^{\vec n} J_3^{\vec n}\frac{S^{\vec n} S^{\vec k}}{S^2}
\crcr
&& - J_1^{\vec n\wedge \vec k}J_3^{\vec n\wedge \vec k} -\frac{1}{S^2} 
 J_1^{ \vec S \wedge(\vec n \wedge \vec k) } 
 J_3^{ \vec S \wedge(\vec n \wedge \vec k) } 
+\vec J_1 \cdot \vec J_3 \frac{(\vec nS^{\vec k}+\vec k S^{\vec n})^2}{S^2} \; ,
\eea
and computing the middle term on the second line taking into account $\vec S  \cdot \vec J_i=0$, we obtain
\bea
&& (\vec J_1\wedge \vec n)\cdot(\vec J_3\wedge \vec n) \Big{(} (\vec n \wedge \vec k)^2 -
4 \frac{ (S^{\vec n} S^{\vec k})^2}{S^2}  \Big{)} 
-4(\vec n\cdot \vec k) J_1^{\vec n} J_3^{\vec n}\frac{S^{\vec n} S^{\vec k}}{S^2}
\crcr
&&+\vec J_1 \cdot \vec J_3 \frac{(\vec nS^{\vec k}+\vec k S^{\vec n})^2- (\vec nS^{\vec k}-\vec k S^{\vec n})^2}{S^2}
\; ,
\eea
hence the real part is 
\bea
(\vec J_1\wedge \vec n)\cdot(\vec J_3\wedge \vec n) \Big{[} (\vec n \wedge \vec k)^2 -4 \frac{ (S^{\vec n} S^{\vec k})^2}{S^2}  + 4  \frac{S^{\vec n} S^{\vec k} (\vec n\cdot \vec k)}{S^2} \Big{]} \; .
\eea

The imaginary part of the numerator (\ref{eq:numera}) writes
\bea
J_1^{\vec S \wedge \vec n \frac{S^{\vec k}}{S^2} + \vec S \wedge \vec k \frac{S^{\vec n}}{S^2} }
J_3^{\vec n \wedge (\vec k \wedge \vec n) + 2 \vec n  \frac{  S^{\vec n} S^{\vec k} }{S^2}  }
-J_1^{\vec n \wedge (\vec k \wedge \vec n) + 2 \vec n  \frac{  S^{\vec n} S^{\vec k} }{S^2}  }
J_3^{\vec S \wedge \vec n \frac{S^{\vec k}}{S^2} + \vec S \wedge \vec k \frac{S^{\vec n}}{S^2} }
\; .
\eea
We start by expressing it as 
\bea
&& -\vec J_1^{\;\wedge \Bigl[\vec n \wedge (\vec k \wedge \vec n) + 2 \vec n  \frac{  S^{\vec n} S^{\vec k} }{S^2}  \Bigr]}
\cdot
\vec J_3^{\;\wedge \Bigl[\vec S \wedge \vec n \frac{S^{\vec k}}{S^2} + \vec S \wedge \vec k \frac{S^{\vec n}}{S^2} \Bigr]}
\crcr
&& +\vec J_3^{\;\wedge \Bigl[\vec n \wedge (\vec k \wedge \vec n) + 2 \vec n  \frac{  S^{\vec n} S^{\vec k} }{S^2}  \Bigr]}
\cdot 
\vec J_1^{\;\wedge \Bigl[\vec S \wedge \vec n \frac{S^{\vec k}}{S^2} + \vec S \wedge \vec k \frac{S^{\vec n}}{S^2} \Bigr]}
\; ,
\eea
as the cross terms in the development of the two scalar products cancel. This computes further to
\bea
&&- \Big{(} \vec n J_1^{\vec k\wedge \vec n} - (\vec k \wedge \vec n)  J_1^{\vec n} +
 2 \vec J_1\wedge \vec n \frac{  S^{\vec n} S^{\vec k} }{S^2}  \Big{)} \cdot
\vec S \Big{(} J_3^{\vec n} \frac{S^{\vec k}}{S^2} +J_3^{\vec k} \frac{S^{\vec n}}{S^2} \Big{)} \crcr
&&+\Big{(} \vec n J_3^{\vec k\wedge \vec n} - (\vec k \wedge \vec n)  J_3^{\vec n} +
 2 \vec J_3\wedge \vec n \frac{  S^{\vec n} S^{\vec k} }{S^2}  \Big{)} \cdot
\vec S \Big{(} J_1^{\vec n} \frac{S^{\vec k}}{S^2} +J_1^{\vec k} \frac{S^{\vec n}}{S^2} \Big{)} \; .
\eea
Grouping together similar terms we get
\bea
&& \frac{S^{\vec n}S^{\vec k}}{S^2} \Big{(}\vec J_3 \wedge \vec J_1 \Big{)}
 \cdot \Big{(}(\vec k\wedge \vec n)\wedge \vec n \Big{)} + 
\frac{(S^{\vec n})^2}{S^2}  \Big{(}\vec J_3 \wedge \vec J_1 \Big{)}
 \cdot \Big{(}(\vec k\wedge \vec n)\wedge \vec k \Big{)} 
\crcr
&&-[(\vec k \wedge \vec n)\cdot \vec S] \frac{S^{\vec n}}{S^2} \Big{(}\vec J_3 \wedge \vec J_1 \Big{)}
\cdot (\vec n\wedge \vec k) \\
&&+2 \frac{S^{\vec n}(S^{\vec k})^2}{S^4}\Big{[} \Big{(}\vec n \wedge(\vec J_3\wedge \vec J_1)
 \Big{)} \wedge \vec n\Big{]}\cdot \vec S 
+2 \frac{(S^{\vec n})^2 S^{\vec k}}{S^4}\Big{[} \Big{(}\vec k \wedge(\vec J_3\wedge \vec J_1)
 \Big{)} \wedge \vec n\Big{]}\cdot \vec S \; .\nonumber
\eea
Recognizing $\vec S = \vec J_3 \wedge \vec J_1  $, the above writes
\bea
&& \frac{S^{\vec n}S^{\vec k}}{S^2} [S^{\vec n} (\vec n\cdot \vec k) - S^{\vec k}]
+\frac{(S^{\vec n})^2}{S^2}  [S^{\vec n}- S^{\vec k} (\vec n\cdot \vec k)] 
+\frac{S^{\vec n}}{S^2} [\vec S \cdot(\vec n\wedge \vec k)]^2 \crcr
&&+2 \frac{S^{\vec n}(S^{\vec k})^2}{S^4} \Big{(} - (S^{\vec n})^2 + S^2\Big{)}
+2 \frac{(S^{\vec n})^2 S^{\vec k}}{S^4} \Big{(} - S^{\vec n}S^{\vec k} + (\vec n\cdot \vec k)S^2\Big{)}
\crcr
&&=S^{\vec n} \Big{[} (\vec n\wedge \vec k)^2 - \frac{(\vec n S^{\vec k} -\vec k S^{\vec n})^2}{S^2}
-\frac{(S^{\vec k})^2}{S^2}+\frac{(S^{\vec n})^2}{S^2} - 4 
\frac{(S^{\vec n} S^{\vec k})^2}{S^4} 
+2 \frac{(S^{\vec k})^2}{S^2} +
2   (\vec n\cdot \vec k)  \frac{S^{\vec n} S^{\vec k}}{S^2}
\Big{]}\crcr
&&=S^{\vec n} \Big{[} (\vec n\wedge \vec k)^2 +4 \frac{S^{\vec n} S^{\vec k}}{S^2} - 
4 \frac{(S^{\vec n} S^{\vec k})^2}{S^4}   \Big{]} \; .
\eea

In conclusion $\imath  \epsilon_1^{-} \psi_1^{-}- \imath  \epsilon_3^{-} \psi_3^{-}$ is 
\bea
 \imath  \epsilon_1^{-} \psi_1^{-}- \imath  \epsilon_3^{-} \psi_3^{-}  = 
\ln \frac{ (\vec n \wedge \vec J_1)\cdot (\vec n \wedge \vec J_3)+ 
 \imath \vec n \cdot (\vec J_3 \wedge \vec J_1)}{\sqrt{ (\vec n\wedge \vec J_1)^2(\vec n\wedge \vec J_3)^2 }  }
 = \imath \Psi^{13}_{\vec n} \; .
\eea

Following similar manipulations we get
\bea
\imath  \epsilon_2^{\pm} \psi_2^{\pm}- \imath  \epsilon_3^{\pm} \psi_3^{\pm} =
\ln \frac{ (\vec n \wedge \vec J_2)\cdot (\vec n \wedge \vec J_3)+ 
 \imath \vec n \cdot (\vec J_3 \wedge \vec J_2)}{\sqrt{ (\vec n\wedge \vec J_2)^2(\vec n\wedge \vec J_3)^2 }  }
 = \imath \Psi^{23}_{\vec n} \; .
\eea

For the angles $\omega_i$, recall that $\omega_{\vec n,\vec k}$ can be written in terms of $\psi_{-\vec k,-\vec n}$.
Note that due to the choice of the determination of the $\sqrt{\Delta_i^+}$ the correct relation is
$\omega^+_{\vec n,\vec k} = -\psi^+_{-\vec k, -\vec n}$ and $\omega^-_{\vec n,\vec k} = \psi^-_{-\vec k, -\vec n}$.
Moreover, as $\Psi_{-\vec k} = -\Psi_{\vec k}$, we conclude
\bea
\imath  \epsilon_1^{\pm} \omega_1^{\pm}- \imath  \epsilon_3^{\pm} \omega_3^{\pm}  =  \pm \imath \Psi^{13}_{\vec k}
\; , \qquad
\imath  \epsilon_2^{\pm} \omega_2^{\pm}- \imath  \epsilon_3^{\pm} \omega_3^{\pm}  =  \pm \imath \Psi^{23}_{\vec k}
\; .
\eea

\section{Boundary terms in the Euler Maclaurin formula}
\label{app:EM}

Using the short hand notation $F(t)$ for $F(J,M,M',t)$, the remainder terms in the EM 
formula write
\bea
- B_1  \bigl[  F(t_{\max}) +F(t_{\min}) \bigr] + \sum_{k}\; \frac{B_{2k}}{(2k)!}\,
 \bigl[ F^{(2k-1)}(t_{\max}) -F^{(2k-1)}(t_{\min}) \bigr] \; .
\label{remain}
\eea
In this section we deal with generic Wigner matrices, that is we consistently assume that $0<\xi^2<1$.
Note that
\bea
 t_{\min}= \max \{0, M - M' \} \; ,\qquad t_{\max}= \min \{ J+M, J-M' \} \; .  
\eea
For simplicity we will detail the diagonal matrix elements $M=M'$. By continuity the region in which 
our results apply extends to some strip $|M-M'|<P$. For such elements $t_{\min}=0$ and, 
for $M>0$, $t_{\max} = J-M$. The Stirling approximations 
become easily upper and lower bounds, at the price of some constants, thus by Appendix \ref{app:stir} we obtain 
\bea\label{eq:bondstir}
\frac{C^{\min}}{J} \sqrt{K(x,x,u)}e^{J \Re f (x,x,u)} <|F(t)| < \frac{C^{\max}}{J} \sqrt{K(x,x,u)}e^{J \Re f (x,x,u)} \; ,
\eea
with 
\bea
&&f(x,x,u) =
-\imath(\alpha+\gamma) x +\imath \pi u + (1-u) \ln \xi^2
+ u \ln (1-\xi^2)  \crcr
&&
+ (1-x) \ln(1-x) +   (1+x) \ln(1+x)  -2 u\ln u 
\crcr
&&-  (1+x-u)\ln(1+x-u) - (1-x-u) \ln(1-x-u) \; .
\eea
and 
\bea
 K(x,u) = \frac{(1-x^2)}{(1+x-u)(1-x-u)u^2  } \; .
\eea

The behavior of the higher derivative terms in the EM formula is governed by 
 $F^{(k)}(t_{\min})$ and $F^{(k)}(t_{\max})$. To see this,
 collect all factors depending on $t$ in $F(t)$ and write
\bea
&&
F(t) = q(J,M) \; p_{J,M}(t) \; ,\crcr
&& p_{J,M}(t) = \frac{e^{A t}}
{\Gamma(J+M-t+1) \Gamma(J-M-t+1)[\Gamma(t+1)] ^2} \; ,\crcr
&& A := \imath[\pi   -2 \ln \xi - 2 \ln \eta].
\label{eq:pjmm}
\eea
Hence $ F^{(k)} = q(J,M) p^{(k)}_{J,M}$, and the first derivatives expresses in terms of
\bea
\frac{d}{dt}p_{J,M}(t) &=& p_{J,M}(t) \Big{\{}A +\frac{\Gamma'(J+M-t+1)}{\Gamma(J+M-t+1)} \crcr
&& + \frac{\Gamma'(J-M-t+1)}{\Gamma(J-M-t+1)} - 2 \frac{\Gamma'(t+1)}{\Gamma(t+1) }
\Big{\}}  \crcr
&=&  p_{J,M}(t)\Big{\{}A +\psi^{(0)}(J+M-t+1) \crcr
&& + \psi^{(0)}(J-M-t+1) - 2 \psi^{(0)}(t+1) \Big{\}}\; ,
\label{eq:derpante}
\eea
with $\psi^{(0)}(t)$ denoting the digamma  function. For integer arguments 
\bea
 \psi^{(0)}(m+1)= -\gamma_0 + \sum_{k=1}^m \frac1k \; ,
\eea
hence $|F'(t)|<C \; \ln J \; |F(t)|$ for some constant $C$.
Higher order derivatives of eq. (\ref{eq:derpante}) write in terms of higher order polygamma functions
 $\psi^{(n)} = d^n \psi^{(0)}/dt^n$. For all $k$, $\psi^{(2k)}(X) \leq \psi^{(0)}(X)$
at large $X$, therefore the $k$'th derivative is dominated by 
\bea
F^{(2k-1)}(t) =  F(t) \left\{ \bigl[A + \sum_{i=1}^4 \pm \psi^{(0)}(X_i) \bigr]^{2k-1}+ \ldots\right\} \; .
\eea
Then $|F^{(k)}|< C (\ln J)^k |F(t)|$ for some constant $C$. 

\begin{figure}
 \centerline{\includegraphics[width=3cm]{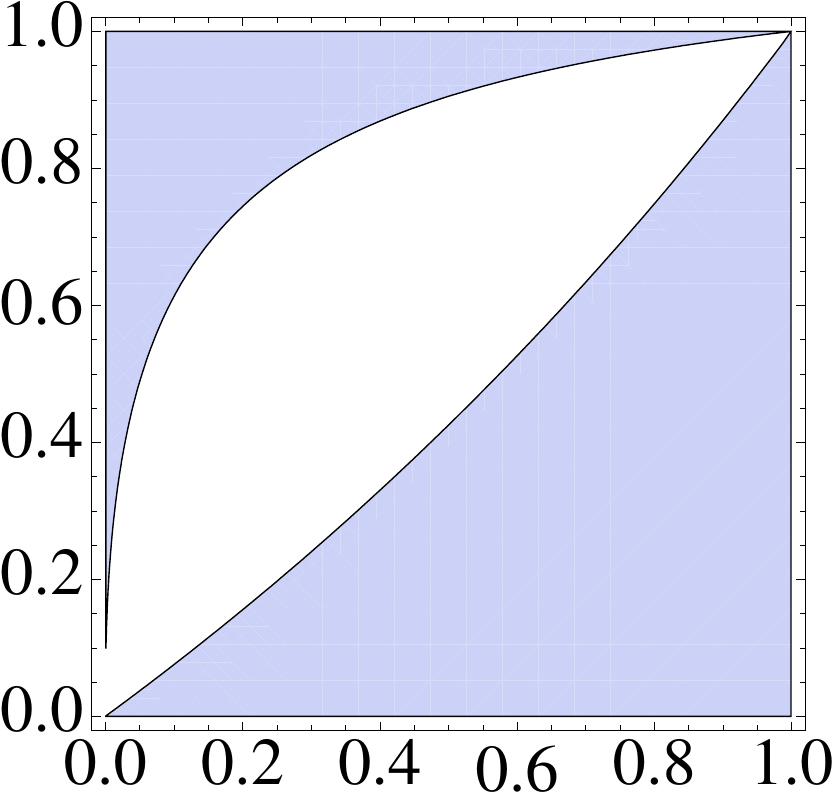}\put(-100,40){$\xi$}
\put(-43,-10){$x$}}
\caption{Shaded region where the EM corrections are exponentially suppressed}
\label{fig:2}
\end{figure}

From eq. (\ref{eq:bondstir}) we conclude that both $|F(t_{\min})|$ and $|F(t_{\max})|$, as well as all their
derivatives are a priori exponentially suppressed in the region where 
$ \Re f(x,y, u_{\min} ) <0 $ and $\Re f(x,y, u_{\max})<0$. As 
\bea
 \Re f(x,x,0) &=&  \ln \xi^2  \nonumber\\
 \Re f(x,x,1-x) &=& x\ln\xi^2 +(1-x) \ln (1-\xi^2) \crcr
&& + (1+x)\ln(1+x)-(1-x)\ln(1-x) -2x\ln (2x) \; ,
\eea
we conclude that the derivative corrections coming from $t_{\min}=0$ are always suppressed term by term. 
However the situation is markedly different for the corrections coming from $t_{\max}=J-M$. At fixed 
$\xi^2$, the corrections are exponentially suppressed for $x$ close enough to either $0$ or $1$, 
but the maximum of $\Re f(x,x,1-x) $, achieved for $x= \frac{\xi}{\sqrt{ 4-3\xi^2  }}$ is
$ \ln \frac{(\xi+\sqrt{4-3\xi^2})^2}{4} >0$, hence there exists some interval in which, term by term,
the derivative terms are bounded from below by an exponential blow up. In this region our EM SPA approximation
should a priori fail (see also figure \ref{fig:2}).

A second set of EM derivative terms come when passing from eq. (\ref{eq:charsum}) to eq. (\ref{eq:char}), involving derivatives
 $\frac{\partial^n}{(\partial_{x})^n} D^{J}_{xJ,xJ}|_{x=\pm 1}$. Using Appendix \ref{app:delta<0}, eq. (\ref{eq:delt<0est})
we note that all these derivatives yield some function times $D^J_{xJ,xJ}$. As 
\bea
&&
D^J_{-J,-J}(g) = \xi^{2J} e^{+\imath(\alpha+\gamma) J}\,, \qquad
D^J_{JJ}(g) = \xi^{2J} e^{-\imath(\alpha+\gamma) J} \;,
\label{dmindmax}
\eea
all such derivative terms are exponentially suppressed for large $J$.

\end{document}